\documentclass[useAMS,usenatbib]{mn2e}

\usepackage{graphicx}
\usepackage{epsfig}
\usepackage{amssymb,amsmath}
\usepackage{multirow}
\usepackage{mathrsfs}
\usepackage{times}
\usepackage{setspace}
\usepackage{color}

\usepackage[dvipsnames]{xcolor}

\usepackage[colorlinks, breaklinks, citecolor=blue]{hyperref}

\newcommand{\calL}{\mathcal{L}}
\newcommand{\calG}{\mathcal{G}}
\newcommand{\calT}{\mathcal{T}}

\newcommand{\phint}{\phi_{\mathrm{nt}}}
\newcommand{\phimp}{\phi_{\mathrm{mp}}}
\newcommand{\Qmin}{Q_{\rm min}}
\newcommand{\rhogmp}{\rho_{\rm g,mp}}

\title[Galactic Discs]{A Unified Model for Galactic Discs: Star Formation, Turbulence Driving, and Mass Transport}

\author[Krumholz et al.]{Mark R. Krumholz$^1$\thanks{mark.krumholz@anu.edu.au},
Blakesley Burkhart$^2$,
John C.~Forbes$^2$, and Roland M.~Crocker$^1$
\\ \\
$^1$ Research School of Astronomy \& Astrophysics, Australian National University, Canberra, ACT 2611, Australia\\
$^2$ Harvard-Smithsonian Center for Astrophysics, 60 Garden St, Cambridge, MA 0213, USA
}

\begin{document}
\maketitle
\label{firstpage}
\begin{abstract}
We introduce a new model for the structure and evolution of the gas in galactic discs. In the model the gas is in vertical pressure and energy balance. Star formation feedback injects energy and momentum, and non-axisymmetric torques prevent the gas from becoming more than marginally gravitationally unstable. From these assumptions we derive the relationship between galaxies' bulk properties (gas surface density, stellar content, and rotation curve) and their star formation rates, gas velocity dispersions, and rates of radial inflow. We show that the turbulence in discs can be powered primarily by star formation feedback, radial transport, or a combination of the two. In contrast to models that omit either radial transport or star formation feedback, the predictions of this model yield excellent agreement with a wide range of observations, including the star formation law measured in both spatially resolved and unresolved data, the correlation between galaxies' star formation rates and velocity dispersions, and observed rates of radial inflow. The agreement holds across a wide range of galaxy mass and type, from local dwarfs to extreme starbursts to high-redshifts discs. We apply the model to galaxies on the star-forming main sequence, and show that it predicts a transition from mostly gravity-driven turbulence at high redshift to star formation-driven turbulence at low redshift. This transition, and the changes in mass transport rates that it produces, naturally explain why galaxy bulges tend to form at high redshift and discs at lower redshift, and why galaxies tend to quench inside-out.
\end{abstract}

\begin{keywords}
galaxies: formation --- galaxies: ISM --- galaxies: star formation --- ISM: kinematics and dynamics --- stars: formation --- turbulence 
\vspace{0.5in}
\end{keywords}

%\clearpage

\section{Introduction}
\label{sec:intro}

\subsection{Observational Background}

Despite their diversity in mass, spatial extent, and stellar and gas content, disc galaxies both in the local and distant Universe show a striking range of regularities. Perhaps the most famous of these is the Kennicutt-Schmidt relation \citep[see reviews by][]{kennicutt98a, kennicutt12a, krumholz14c}, the observed correlation between the rate at which galaxies form stars and a combination of their gas content and their dynamical times. The rate of star formation implied by this relation is remarkably small: on average, galaxies turn only $\sim 1\%$ of their gas into stars per dynamical time of the gas \citep{zuckerman74a}. This correlation between gas content and star formation, and the remarkably low efficiency of star formation that it implies, was first observed on galactic scales and in the local Universe. However, subsequent work has shown that it continues to hold even at high redshift \citep[e.g.,][]{bouche07a, daddi08a, daddi10a, daddi10b, genzel10a, tacconi13a}, and on $\sim 1$ kpc scales in the local Universe \citep{kennicutt07a, bigiel08a, leroy08a, leroy13a, liu11a, momose13a}.

Indeed, the correlation and inefficiency extend down to even $\sim 1$ pc scales. There are a number of lines of evidence in favour of this conclusion, including direct star counts in star-forming clouds near the Sun \citep{lada10a, heiderman10a, krumholz12a, evans14a, salim15a, heyer16a}, correlations between gas and indirect star formation tracers such as recombination lines to larger distances in the Milky Way \citep{vutisalchavakul16a}, and correlations between star formation and tracers of dense gas in both Galactic and extragalactic systems \citep{krumholz07e, garcia-burillo12a, usero15a}.\footnote{This conclusion has recently been questioned by \citet{lee16a}, but we argue in this paper that this is likely an artefact of their methodology, which differs from that of all the other authors. See below for details.}

A second regularity and noted galactic-scale correlation concerns the velocity dispersions of the gas in galaxies. In both local and high redshift galaxies, this gas invariably displays superthermal linewidths indicative of transsonic or supersonic motion \citep[and references therein]{glazebrook13a}. This is true regardless of whether these motions are traced using the 21 cm line of H~\textsc{i} \citep{van-zee99a, petric07a, tamburro09a, burkhart10a, ianjamasimanana12a, ianjamasimanana15a, stilp13a, chepurnov15a}, the low$-J$ lines of CO \citep{caldu-primo13a, caldu-primo15a, meidt13a, pety13a}, or the recombination lines of ionised gas \citep{cresci09a, lehnert09a, lehnert13a, green10a, green14a, le-tiran11a, swinbank12a, arribas14a, genzel14a, moiseev15a}. Observed linewidths are relatively independent of radius within a given galaxy, but vary significantly from galaxy to galaxy in a way that is well-correlated with galaxies' rates of star formation. Galaxies with star formation rates below $\sim 1$ $M_\odot$ yr$^{-1}$, typical of the local Universe \citep{kennicutt12a}, all have roughly the same velocity dispersion of $\approx 10$ km s$^{-1}$. However, at the higher star formation rates found both in local starbursts and in main sequence star-forming galaxies at higher redshift, velocity dispersions increase roughly linearly, $\sigma \propto \dot{M}_*$ \citep{krumholz16a}, although with substantial scatter and subsidiary dependencies on quantities such as the galaxies' gas fractions, sizes, and rotational velocities.

These velocity dispersions feed naturally into a third observed correlation, which is that galaxy discs tend to be in a state of marginal gravitational stability. The gravitational stability of a disc can be characterised by the \citet{toomre64a} $Q$ parameter, defined by $Q \approx \kappa\sigma/\pi G \Sigma$, where $\kappa$ is the epicyclic frequency of the galaxy's rotation, $\sigma$ is the velocity dispersion, and $\Sigma$ is the surface mass density. Observed disc galaxies in both the local universe and at high redshift tend to have $Q\approx 1$ throughout their discs \citep[e.g.,][]{martin02a, genzel10a, meurer13a, romeo13a, romeo17a}.

A fourth and final observed correlation relates to the spatial distribution of gas and star formation in galaxy disks. Star formation correlates with molecular gas rather than total gas, and the H$_2$-rich regions of galaxies are preferentially located in their centres. Consequently, the scale length of the star formation is comparable to the stellar scale length, $\approx 2-4$ kpc, and a factor of $2-3$ smaller than the neutral gas scale length  \citep{regan01a, leroy08a, schruba11a, bigiel12a}. Within the molecule-dominated region, the gas depletion time is $\sim 1-2$ Gyr \citep{bigiel08a, leroy13a}, much less than a Hubble time. The fact that star formation has not ceased in the centres of all galaxy discs implies either that we live at a special time when all local disc centres are about to quench, or that there is an ongoing gas supply to fuel star formation. Direct accretion of cold gas from the intergalactic medium \citep[e.g.,][]{keres05a, dekel06a, dekel09b, wetzel15a} and condensation from hot halos in low redshift galaxies \citep{marinacci10a, joung12a, fraternali13a, hobbs13a}, supplemented by mass returned by stellar evolution \citep{leitner11a}, likely provide a sufficient mass supply for star formation. However, they do not naturally provide it at the small galactocentric radii where star formation takes place. Accretion from a hot corona is predicted to deliver most of its mass at radii of $\sim 3-4$ stellar scale lengths \citep[e.g.,][]{marasco12a}, and, at least at high redshift, cold accretion tends to join the disc at radii of $\sim 0.1 - 0.3$ virial radii, which is $\sim 10$ times the stellar scale length \citep{danovich15a}, though there are exceptions associated with loss of angular momentum by counter-rotating streams and major mergers, which tend to trigger ``compaction" events \citep{zolotov15a, tacchella16a, tacchella16b}. Preventing quenching requires this gas then flow radially inward. Such flows have recently been detected directly in a number of nearby galaxy discs \citep{schmidt16a}.

\subsection{Theoretical Background}
\label{ssec:theory}

Any successful theory of the structure and evolution of disc galaxies ought to be able to explain all of these observed regularities, but at present no such theory is available. This is at least in part because theoretical modelling has tended to focus on one or two of the observed correlations, without attempting to unify all of them into a single, coherent picture.

Several authors have attempted to develop theories that link the problems of velocity dispersion, marginal stability, and star formation fuelling \citep[e.g.,][]{bournaud07a, bournaud09a, agertz09b, dekel09a, ceverino10a, krumholz10c, vollmer11a, cacciato12a, forbes12a, forbes14a, goldbaum15a, goldbaum16a}. The central premise in these models is that gravitational instability produces torques that both move mass inward and drive turbulence, simultaneously regulating galaxies to $Q\approx 1$, producing supersonic velocity dispersions, and fuelling star formation. Models in this class naturally explain why $Q\approx 1$, why star formation is not quenched in modern galaxy centres, and why high redshift galaxies have high velocity dispersions. If one couples them to an empirically-determined star formation relation, they can also  do a reasonable job of explaining both galaxy-scale star formation laws and the high star formation - velocity dispersion portion of the $\sigma-\dot{M}_*$ correlation \citep{zheng13a, krumholz16a, wong16a}.

However, these models do not naturally explain the minimum velocity dispersion to which galaxy discs seem to settle at $z\approx 0$. Even in quiescent galaxies similar to the Milky Way, observed ISM velocity dispersions are $\approx 10$ km s$^{-1}$ \citep[e.g.,][]{ianjamasimanana12a}, corresponding to bulk motions at a Mach number $\sim 1$ for gas at the typical warm neutral medium temperature of $\approx 7000$ K \citep{wolfire03a}. Some energy input is required to maintain transsonic flows of this sort, and models based purely on gravitational instability-driven torques do not naturally produce such an input in quiescent discs. Because such models do not naturally make any predictions about star formation rates on either large or small scales, they also do not explain the physical origins of the star formation law. More generally these models usually do not include any specific treatment of star formation feedback or its coupling to the interstellar medium, an obvious omission.

Other authors have instead chosen to focus on the observed correlation between star formation and gas. Some authors have attempted to derive this correlation using a ``bottom up" approach, whereby one begins by attempting to explain the inefficiency of star formation on small scales, and then builds a galaxy-scale star formation relation as the sum of small-scale relations \citep{krumholz05c, krumholz09b, padoan12a, federrath12a, federrath13a, krumholz13c, federrath15b, burkhart18a}. These small-scale relations, while theoretically-motivated, can be checked directly against numerical simulations of self-gravitating turbulence, and the agreement is generally good \citep[e.g.,][]{padoan11a, federrath12a, padoan14a}. This approach allows one to explain the star formation rate on both small and large scales, and naturally incorporates star formation feedback on small scales. Furthermore, if these models are supplemented by chemical models that capture the transition between the warm, H~\textsc{i} and cold, H$_2$ phases of the ISM \citep{krumholz09b, krumholz13c}, they also correctly capture the observed dependence of the star formation rate on the chemical phase and metallicity of the ISM \citep{bolatto11a, wong13a, shi14a, filho16a, jameson16a, rafelski16a}. On the other hand, these models are generally silent on the question of galaxies' velocity dispersions, gravitational stability, or long term fuelling.

Conversely, some authors have attempted to derive the star formation rate and velocity dispersion using a ``top down" method, the fundamental assumption of which is that the star formation rate is set by considerations of force and energy balance within a galactic disc \citep[e.g.,][]{thompson05a, ostriker10a, ostriker11a, hopkins11a, faucher-giguere13a, hayward17a}. In these models, one considers a disc of a prescribed gas content and gravitational potential, and asks what star formation rate is required for star formation feedback to be vigorous enough to keep the disc in vertical pressure balance and energy balance. This approach has the advantage that it is rooted in simple physical considerations that must hold at some level, and it is the first step in the approach that we shall pursue in this paper. Moreover, it enables one to make predictions that link star formation, velocity dispersion, and Toomre stability, and thus unify three of the observed correlations discussed above. 

However, top-down models that work solely based on the balance between feedback, vertical gravity, and dissipation have proven difficult to make work in practice. First of all, unless one posits a source of star formation feedback for which the momentum injected per star formed increases with gas surface density \citep[e.g., as trapped infrared radiation pressure does in the model of][]{thompson05a}, the natural prediction of these models is that the star formation rate per unit area should rise as the square of the gas surface density (e.g., equation 13 of \citealt{ostriker11a} or equation 18 of \citealt{faucher-giguere13a}). The predicted correlation $\dot{\Sigma}_* \propto \Sigma^p$ with $p \approx 2$ is substantially steeper than the observed correlation, which ranges between $p\approx 1$ in spatially resolved patches of local galaxies to $p\approx 1.5$ for rapidly star-forming galaxies as a whole.\footnote{\citet{narayanan12a} and \citet{faucher-giguere13a} argue that one can steepen the relation and increase the value of $p$ by adopting a CO to H$_2$ conversion factor that scales strongly with galaxy star formation rate. However, even adopting such a scaling, fits to the more recent and larger data sets favour $p \approx 1.7$ rather than $2$ \citep[c.f.~figure 3 of][]{thompson16a}, and recent dust-based measurements of gas content that are independent of CO suggest that even this is too steep \citep{genzel15a}.}

Second, because these models compute the star formation rate from the weight of the ISM, they naturally predict that the star formation rate at a given surface density is independent of the metallicity or chemistry of the ISM, since these factors do not alter the weight. They can be reconciled with the strong observational evidence that metallicity and chemical phase do affect the rate of star formation only by positing  that the efficiency of star formation feedback is metallicity-dependent. For example, the model of \citet{ostriker10a} predicted that the regions of comparable gas surface density in the Small Magellanic Cloud and the Milky Way should form stars at nearly equal rates. \citet{bolatto11a} found that this prediction was incorrect, and proposed a modification to the theory in which the efficiency of photoelectric heating scales inversely with metallicity, and thus stars pressurise the ISM more efficiently in low-metallicity galaxies. While this does fix agreement with the observations, \citet{krumholz13c} points out that the physical mechanism proposed by \citet{bolatto11a} to produce the metallicity dependence is not correct, and, more generally, that there is no good reason to expect that feedback efficiencies will depend on metallicity in the ways required to explain the observations. In particular, supernovae are thought to be the dominant feedback mechanism in most galaxies, and supernova momentum injection is nearly independent of metallicity \citep[e.g.,][]{thornton98a, martizzi15a, gentry17a}.

Third, these models do not naturally predict either the sub-galactic star formation law or the gravitational stability parameter, forcing one to adopt one or the other based on empirical observations. If one adopts the observed sub-galactic star formation rate \citep[e.g.,][]{ostriker11a}, then, as we shall show below, one predicts velocity dispersions and Toomre $Q$ parameters sharply at odds with what is observed. Conversely, one can posit that star formation rates are very sensitive to the Toomre $Q$ parameter, so that the star formation rate self-adjusts to maintain $Q\approx 1$ \citep[e.g.,][]{faucher-giguere13a, hayward17a}. By construction this produces the correct Toomre $Q$, but it still fails to reproduce the observed $\sigma-\dot{M}_*$ correlation (because the predicted star formation law is too steep -- \citealt{krumholz16a}), and it also predicts that star formation on small scales is very efficient in high surface density galaxies, contrary to observations. Just to give one example of this difficulty: if star formation efficiencies on small scales were higher in high surface density galaxies, then the ratio of infrared emission (a star formation tracer) to HCN luminosity (a tracer of dense gas on small scales) should increase with star formation rate, whereas the observed trend is the opposite \citep{garcia-burillo12a, usero15a}. This approach also runs into observational difficulty with its central assumption that galaxies' star formation rates are very sensitive to the value of the Toomre $Q$ parameter; observations strongly disfavour any such correlation \citep{leroy08a}. Instead, both observations and simulations \citep{agertz09a, goldbaum15a, goldbaum16a} seem to suggest that the response of a disc to a drop in Toomre $Q$ is that the disc becomes non-axisymmetric and moves mass inwards, rather than that its star formation rate dramatically increases.

\subsection{This Work and Its Motivation}

Our goal in this work is to unify models of galactic discs that focus on transport, star formation fuelling, and gravitational instability with those that focus on the energy and momentum balance of star formation feedback. We show below that this approach remedies many of the observational problems we have identified with the various theories that have been proposed to date. However, the need for such a synthesis can be driven home simply by more basic consideration of the observations and their energetic implications.

The turbulent energy per unit area contained in a galactic disc of gas surface density $\Sigma_{\rm g}$ and velocity dispersion $\sigma_{\rm g}$ is
\begin{equation}
\left(\frac{dE}{dA}\right)_{\rm turb} \approx \frac{3}{2}\Sigma_{\rm g} \sigma_{\rm g}^2 = 3.1\times 10^9\, \Sigma_{\rm g,10} \sigma_{\rm g,10}^2\mbox{ erg cm}^{-2},
\end{equation}
where $\Sigma_{\rm g,10} = \Sigma_{\rm g}/10$ $M_\odot$ pc$^{-2}$ and $\sigma_{\rm g,10} = \sigma/10$ km s$^{-1}$; the scaling factors are typical values at the Solar Circle in the Milky Way. The energy should dissipate due to decay of turbulence over a timescale comparable to the crossing time, but, in a disc with $Q\approx 1$, this timescale is comparable to the galactic dynamical time $t_{\rm dyn} = r/v_\phi$, where $r$ is the galactocentric radius and $v_\phi$ is the rotation velocity. It is therefore of interest to consider possible sources of power that are capable of delivering this amount of energy per unit area over a timescale $t_{\rm dyn}$.

As noted above, \citet{schmidt16a} directly detect flows of mass radially inwards through the discs of local spiral galaxies with mass fluxes $\dot{M}_{\rm in} \sim 1$ $M_\odot$ yr$^{-1}$. These observations are difficult due to the near cancellation of inflow- and outflow-rates around spiral arms and in outer regions where galaxies become significantly lopsided. At a minimum, the magnitude of the inflow should be regarded as significantly uncertain. However, we note that inflows rates of roughly this size must be ubiquitous to explain star formation fuelling. In a galaxy with a flat rotation curve, the amount of gravitational potential energy per unit area per unit time released by this flow of mass down the potential well is
\begin{equation}
\frac{d^2 E}{dt\, dA} \approx \frac{\dot{M}_{\rm in} v_\phi^2}{2\pi r^2},
\end{equation}
so over a galactic dynamical time the flow delivers an energy per unit area
\begin{equation}
\left(\frac{dE}{dA}\right)_{\rm inflow} \approx \frac{\dot{M}_{\rm in} v_\phi}{2\pi r} = 6.5\times 10^9\, \dot{M}_{\rm in,1} v_{\phi,200} r_{10}^{-1}\mbox{ erg cm}^{-2},
\end{equation}
where $\dot{M}_{\rm in, 1} = \dot{M}/1$ $M_\odot$ yr$^{-1}$, $v_{\phi,200} = v_\phi/200$ km s$^{-1}$, and $r_{10} = r/10$ kpc.

In comparison, star formation feedback is expected to inject energy at a rate per unit area 
\begin{equation}
\frac{d^2 E}{dt\, dA} \approx \dot{\Sigma}_* \left\langle \frac{p_*}{m_*}\right\rangle \sigma_{\rm g},
\end{equation}
where $\dot{\Sigma}_*$ is the star formation rate per unit area and $\langle p_*/m_*\rangle$ is the terminal momentum per unit mass delivered by star formation feedback. (We give a detailed explanation for the origin of this expression below, but intuitively it results simply from the assumption that motions driven by stellar feedback break up and add their energy to the turbulent background once their expansion velocities become comparable to the overall velocity dispersion, so the energy added per ``injection event" is of order the momentum injected times the velocity dispersion.)  Simulations suggest the momentum per unit mass is $\langle p_*/m_*\rangle \approx$ 3000 km s$^{-1}$ for single supernovae \citep{cioffi88a, thornton98a, martizzi15a, kim15a, walch15b, gentry17a, kim17a}. Over a galactic dynamical time, and scaling to Solar Circle values again,
\begin{eqnarray}
\left(\frac{dE}{dA}\right)_{\rm sf} & \approx & \dot{\Sigma}_* \left\langle \frac{p_*}{m_*}\right\rangle \sigma_{\rm g} \frac{r}{v_\phi}
\nonumber \\
& = & 3.1\times 10^9\, \dot{\Sigma}_{*,-3} \sigma_{\rm g,10} r_{10} v_{\phi,200}^{-1}\mbox{ erg cm}^{-2},
\end{eqnarray}
where $\dot{\Sigma}_{*,-3} = \dot{\Sigma}_* / 10^{-3}$ $M_\odot$ pc$^{-2}$ Myr$^{-1}$.

The implication of this calculation is that, at least at the order of magnitude level, inflow and star formation feedback are comparably important energetically in the Solar neighbourhood, and that both are capable of supplying enough energy to replenish the turbulence in the ISM over a galactic dynamical time. Moreover, if we were to repeat this calculation for other types of galaxies we might well get quite different results. The ratio of $(dE/dA)_{\rm inflow}$ to $(dE/dA)_{\rm sf}$ scales as $(\dot{M}_{\rm in}/\dot{M}_*) (v_\phi^2/\sigma_{\rm g})$, where $\dot{M}_*$ is the total star formation rate. We do not have direct measurements of $\dot{M}_{\rm in}$ except in local spirals, but assuming that $\dot{M}_{\rm in}/\dot{M}_* \sim 1$, as would be required to explain star formation fuelling and as is observed locally, star formation should be energetically dominant in galaxies with smaller $v_\phi$ (for example local dwarfs), while inflow should dominate those with larger $\sigma_g$ (for example high-$z$ galaxies). Clearly it is not reasonable to ignore either star formation feedback or inflows in building a model of galaxy discs, as has been the practice for most work up to this point. 

Below we build a minimal unified model that combines both of these processes. We show that, while simple, this model is far more successful than either feedback-only or inflow-only models at explaining the observed correlations obeyed by galaxy discs. We derive the model in \autoref{sec:model}, and compare it to a variety of observations in \autoref{sec:observations}. We discuss the implications of our findings for galaxy formation in \autoref{sec:discussion}, and conclude in \autoref{sec:summary}.

\section{Model}
\label{sec:model}

\begin{table*}
\begin{tabular}{llll}
\hline
Symbol & Fiducial Value & Meaning & Defining equation \\
\hline
\multicolumn{4}{c}{Inputs to model} \\
\hline
$\Sigma_{\rm g}$ & - & Gas surface density & - \\
$\Sigma_*$ & - & Stellar surface density & - \\
$\sigma_{\rm g}$ & - & Gas velocity dispersion (total thermal plus non-thermal) & - \\
$\sigma_*$ & - & Stellar velocity dispersion & - \\
$\rho_{\rm d}$ & - & Dark matter density & - \\
$v_\phi$ & - & Galaxy rotation curve velocity & - \\
$\Omega$ & - & Galaxy angular velocity & - \\
$t_{\rm orb}$ & - & Galaxy orbital period, $t_{\rm orb} = 2\pi/\Omega$ & - \\
$\beta$ & 0 & Rotation curve index, $\beta = d\ln v_\phi d\ln r$ & - \\
$f_{g,Q}$ & 0.5 & Fractional contribution of gas to $Q$ & \ref{eq:fgQ} \\
$f_{g,P}$ & 0.5 & Fractional contribution of gas self-gravity to midplane pressure & \ref{eq:fgP} \\
$f_{\rm sf}$ & - & Fraction of ISM in star-forming molecular phase & \ref{eq:sfr} \\
\hline
\multicolumn{4}{c}{Physics parameters} \\
\hline
$Q_{\rm min}$ & 1 & Minimum possible disc stability parameter & \ref{eq:Qdef} \\
$\phimp$ & 1.4 & Ratio of total pressure to turbulent pressure at midplane & \ref{eq:phimp} \\
$\eta$ & 1.5 & Scaling factor for turbulent dissipation rate & \ref{eq:loss_rate} \\
$\phi_{\rm Q}$ & 2 & One plus ratio of gas to stellar $Q$ & \ref{eq:phiQ} \\
$\phi_{\rm nt}$ & 1 & Fraction of velocity dispersion that is non-thermal & \ref{eq:phith} \\
$\epsilon_{\rm ff}$ & 0.015 & Star formation efficiency per free-fall time & \ref{eq:sfr} \\
$t_{\rm sf,max}$ & 2 Gyr & Maximum star formation timescale & \ref{eq:sfr1} \\
$\phi_a$ & 2 & Offset between resolved and unresolved star formation law normalisations  & \ref{eq:sfr_averaged} \\
\hline
\multicolumn{4}{c}{Model outputs} \\
\hline
$\rho_{\rm min}$ & - & Minimum midplane density required to produce rotation curve & \ref{eq:rho_min} \\
$t_{\rm orb,T}$ & - & Orbital period at which galaxies switch from GMC to Toomre regime & \ref{eq:ttoomre} \\
$\sigma_{\rm sf}$ & - & Gas velocity dispersion that can be sustained by star formation alone & \ref{eq:sigma_sf} \\
$\Sigma_{\rm sf}$ & - & Gas surface density below which star formation alone can sustain turbulence &
\ref{eq:Sigma_sf} \\
$\dot{M}_{\rm ss}$ & - & Steady-state mass inflow rate through the disc & \ref{eq:mdot_steady} \\
\hline
\end{tabular}
\caption{
\label{tab:quantities}
Symbol definitions. The fiducial value listed is the value used in numerical evaluations and plots unless otherwise stated.
}
\end{table*}

In this section we develop a model for a galactic disc in both vertical hydrostatic and energy equilibrium, where the sources of energy input include both star formation feedback and gravitational potential energy released by inward flow of gas through the disc. A central premise of our model is that the gas is dynamically important and capable of adjusting its inflow rate to maintain marginal stability, rather than simply acting as a passive tracer whose transport rate is dictated by the stellar potential independent of the dynamical state of the gas. This premise likely fails in regions where the gas contributes a negligible mass fraction even at the midplane, for example, the central $\sim 3$ kpc of the Milky Way where the Galactic bar dominates the dynamics \citep[e.g.,][]{binney91a}. We argue in \autoref{ssec:limitations} that the vast majority of the interstellar medium by both mass and star formation rate is not found in such regions, so that our model is applicable to the bulk of the ISM and star formation in the Universe. For now, however, we simply take as given that the transport rate is not dictated by a stellar bar or similar structures, but is able to self-adjust.

For convenience we summarise all the quantities used in our model in \autoref{tab:quantities}. We treat our model galaxy as a thin, axisymmetric disc characterised at every radius $r$ by a total gas surface density $\Sigma_{\rm g}$ and 1D gas velocity dispersion $\sigma_{\rm g}$. In addition to gas, the disc contains stars and dark matter. The dark matter has a density $\rho_{\rm d}$, and we assume that its distribution is close to spherical. If there is a spheroidal stellar distribution, we also include its density in $\rho_{\rm d}$. Other stars are in a disc, characterised by a surface density $\Sigma_*$ and a 1D velocity dispersion $\sigma_*$. For simplicity we assume that both $\sigma_{\rm g}$ and $\sigma_*$ are isotropic. In real stellar discs at $z\approx 0$ this assumption fails at the factor of $\sim 2$ level.

The gas and stars orbit within a steady gravitational potential, which we characterise by the velocity $v_\phi$ required for material in orbit to be in balance between centrifugal and gravitational forces in the co-rotating frame. The rotation curve has an index $\beta = d\ln v_\phi /d\ln r$, the angular velocity at radius $r$ is $\Omega = v_\phi/r$, and the orbital period is $t_{\rm orb} = 2 \pi r/v_\phi$.

We provide the source code to perform the computations involved in the model, and produce all the plots included in the paper, at \url{https://bitbucket.org/krumholz/kbfc17}.

\subsection{Gravitational Instability}

A central ansatz of our model, following \citet{krumholz10c}, \citet{cacciato12a}, \citet{forbes12a}, and \citet{forbes14a}, is that gravitational instability-driven transport will prevent the disc from ever becoming more than marginally gravitationally unstable. If the disc begins to become unstable, the instability will break axisymmetry and the subsequent torques will drive mass inward until marginal stability is restored. We therefore begin by expressing this condition. Modern treatments of gravitational instability include the effects of multiple stellar populations as well as gas, along with the effects of finite thickness and the dissipative nature of gas \citep{rafikov01a, romeo10a, romeo11a, elmegreen11a, hoffmann12a, romeo13a}. In this work we use the simple approximation given by \citet{romeo13a}, 
\begin{equation}
\label{eq:Qdef}
Q \approx \left(Q_{\rm g}^{-1} + \frac{2\sigma_{\rm g}\sigma_*}{\sigma_{\rm g}^2+\sigma_*^2} Q_*^{-1}\right)^{-1}
\end{equation}
where
\begin{equation}
\label{eq:Qgas}
Q_{\rm g} = \frac{\kappa\sigma_{\rm g}}{\pi G \Sigma_{\rm g}}
\end{equation}
and similarly for $Q_*$. Here $\kappa = \sqrt{2(\beta+1)}\Omega$ is the epicyclic frequency. This expression is valid as long as $Q_{\rm g} < Q_*$, the quasi-spherical dark matter halo contributes negligibly to the gravitational stability or instability of the system (i.e., $Q_{\rm d} \gg Q_*$, where $Q_{\rm d}$ is the dark matter $Q$), and the ratio of vertical to radial velocity dispersions for the gas and stars is $\gtrsim 0.5$. The latter two conditions hold broadly across all the galaxies we shall consider; the first requires a bit more discussion, which we defer to the end of this section. For convenience, we can rewrite \autoref{eq:Qdef} as
\begin{equation}
\label{eq:Q}
Q = f_{g,Q} Q_{\rm g} 
\end{equation}
where
\begin{equation}
\label{eq:fgQ}
f_{g,Q} \equiv \frac{\Sigma_{\rm g}}{\Sigma_{\rm g} + [2\sigma_{\rm g}^2/(\sigma_{\rm g}^2+\sigma_*^2)] \Sigma_*}.
\end{equation}
The quantity $f_{g,Q}$ can be thought of as defining the effective gas fraction in the disc for the purposes of computing gravitational stability. It clearly behaves as we intuitively expect, in that $f_{g,Q} \rightarrow 1$ for $\Sigma_{\rm g} \gg \Sigma_*$, and $f_{g,Q} \rightarrow 0$ for $\Sigma_{\rm g} \ll \Sigma_*$. In the Solar neighbourhood, which has gas properties $\Sigma_{\rm g} \approx 14$ $M_\odot$ pc$^{-2}$ \citet{mckee15a}, $\sigma_{\rm g} \approx 7$ km s$^{-1}$ \citep{kalberla09a} and stellar properties $\Sigma_* \approx 33$ $M_\odot$ pc$^{-2}$ and $\sigma_* \approx 16$ km s$^{-1}$ \citep{mckee15a}, we have $Q_* \approx Q_{\rm g} \approx 1.5$ and $f_{g,Q} \approx 0.6$.

The condition for stability is that $Q$ be larger than a value $Q_{\rm min}$ of order unity that depends on the thickness of the disc (thicker discs can be stable at lower $Q$) and the gas equation of state (more dissipative equations require higher $Q$ for stability). As a fiducial value we shall adopt $Q_{\rm min} = 1$, which is appropriate for discs that are relatively quiescent. There is some evidence from cosmological simulations that instability can set in at slightly higher $Q \sim 2 - 3$ in the perturbed discs where a greater fraction of the turbulence is in compressive modes that do not support the gas \citep{inoue16a}, but since this is only a factor of $\sim 2$ level effect and only then in some of the galaxies with which we are concerned, we will neglect this complication.

The case $Q_* < Q_{\rm g}$, where stars rather than gas are the most unstable component, requires a bit more attention. Due to the fact that gas is dissipational and thus usually has a lower velocity dispersion than stars, it tends to be the most unstable component in any gas-rich system. Thus we expect $Q_* > Q_{\rm g}$ to hold in local dwarfs and lower-mass spirals, all star-forming galaxies at high-redshift, and in all mergers and starbursts. However, massive local spirals like the Milky Way are sufficiently gas poor ($f_{\rm gas} \sim 10-20\%$ -- \citealt{saintonge11a}) that for the most part they have $Q_* < Q_{\rm g}$:  \citet{romeo17a} find $Q_{\rm g}/Q_* \approx 0.5-10$ for the HERACLES / THINGS sample, with the bulk of the data at $Q_{\rm g}/Q_* \approx 3$. Our expression for $Q$ (\autoref{eq:Qdef}) assumes $Q_{\rm g} < Q_*$, but the equivalent expression for $Q_{\rm g} > Q_*$ \citep{romeo13a} differs only slightly when $Q_{\rm g}$ and $Q_*$ are within a factor of a few of one another. Quantitatively, using the Solar neighbourhood velocity dispersions quoted above ($\sigma_{\rm g} \approx 7$ km s$^{-1}$, $\sigma_* \approx 16$ km s$^{-1}$), the error produced by using \autoref{eq:Qdef} is 10\% for $Q_{\rm g}/Q_* = 3$, and 17\% for $Q_{\rm g}/Q_* = 10$. This is well below the factor of $\approx 2$ uncertainty in $Q_{\rm min}$, so for simplicity we simply use \autoref{eq:Qdef} in all cases, rather than using a different form for large local spirals than for all the other types of galaxies we will consider. One might also worry that, in the $Q_* < Q_{\rm g}$ regime, gravitational instabilities in the stars might not induce perturbations in the gas capable of driving transport. However, \citet{romeo17a} find that the local spirals with $Q_* < Q_{\rm g}$ are also in the regime where perturbations in the gas and stars are strongly coupled (e.g., see their Figure 5), so this is not a concern.

\subsection{Vertical Force Balance}

A second ansatz of our model, following a number of authors \citep[e.g.,][]{boulares90a, piontek07a, koyama09a, ostriker10a} is that the gas is in vertical hydrostatic equilibrium. The spatially-averaged momentum equation for a time-steady isothermal gas reads (\citealt{krumholz17b}, equation 10.9; also see \citealt{kim15b})
\begin{equation}
\label{eq:vert_hydrostatic}
\frac{\partial}{\partial z} \left\langle \rho_{\rm g} \left(\sigma_{\rm th}^2 + v_z^2 + v_A^2\right)\right\rangle - \frac{\partial}{\partial z}\left\langle\frac{B_z^2}{4\pi}\right\rangle -\left\langle\rho_{\rm g} g_z\right\rangle = 0
\end{equation}
where $\rho_{\rm g}$ is the gas density, $\sigma_{\rm th}$ is the gas thermal velocity dispersion, $v_z$ is the vertical velocity, $v_A$ is the gas Alfv\'en speed, $B_z$ is the $z$ component of the magnetic field, $g_z$ is the vertical gravitational acceleration, and we have oriented our coordinate system so the disc midplane lies in the $xy$ plane; the angle brackets denote averaging over the area of the disc, where the area considered is small compared to the disc scale length, but large compared to the size of an individual molecular cloud of star-forming complex. The first term represents the force exerted by the gradient in thermal, turbulent, and magnetic pressure, the second represents the force due to magnetic tension, and the third represents the force due to gravity. Magnetic tension tends to be subdominant except for unusual, artificially-constructed magnetic field configurations, and thus we can generally drop the second term. This expression omits the contribution from cosmic ray pressure, but this is likely comparable to magnetic pressure in importance \citep[e.g.,][]{boulares90a}.

Integrating \autoref{eq:vert_hydrostatic} from $z=0$ to $\infty$, and assuming that $\rho_{\rm g}\rightarrow 0$ and the Alfv\'en speed remains finite as $z\rightarrow\infty$, we have
\begin{equation}
\label{eq:vert_hydrostatic1}
\rhogmp \left(\sigma_{\rm g}^2 + v_A^2\right)_{\rm mp} = -\int_0^\infty \langle \rho_{\rm g} g_z\rangle \, dz,
\end{equation}
where the subscript mp indicates that a quantity is to be evaluated at the disc midplane, and where we have dropped the angle brackets and implicitly understand that midplane terms represent area averages over the midplane; in writing this expression, we have relied on our assumption that the gas velocity dispersion is isotropic, so $\langle \rho_{\rm g} v_z^2\rangle = \rhogmp (\sigma_{\rm g}^2 - \sigma_{\rm th}^2)$. We write the left hand side as
\begin{equation}
\label{eq:phimp}
\rhogmp \left(\sigma_{\rm g}^2 + v_A^2\right)_{\rm mp} \equiv \phimp \rhogmp \sigma_{\rm g}^2,
\end{equation}
where $\phimp$ is a factor that represents the factor by which the midplane pressure exceeds that due to turbulent plus thermal pressure alone, due to magnetic and cosmic ray pressure. Equipartition between magnetic and kinetic degrees of freedom in the directions transverse to the field corresponds to an Alfv\'en Mach number of $2/3$, which is $\phimp = 1.4$ assuming that thermal pressure is unimportant compared to turbulent pressure. A cosmic ray pressure comparable to the magnetic pressure would increase this to $\phimp \approx 2$. On the other hand, if thermal pressure is non-negligible, for example in modern dwarf galaxies, then kinetic-magnetic equipartition implies $\phimp$ closer to unity, since the gas reaches equipartition only between the non-thermal motions and the magnetic field. The differences between $\phimp = 2$ and $\phimp=1$ are small enough that we will not worry about it, and we will simply use $\phimp=1.4$ as our fiducial value.

The term on the right hand side of \autoref{eq:vert_hydrostatic1} depends on the distribution of gas, stars, and dark matter, since each of these components contributes to $g_z$. To parameterise this dependence, note that the potential $\psi$ obeys the Poisson equation, which in cylindrical coordinates (assuming symmetry in the azimuthal direction) reads
\begin{equation}
\frac{1}{r}\frac{\partial}{\partial r}\left(r \frac{\partial \psi}{\partial r}\right) + \frac{\partial^2 \psi}{\partial z^2} = 4 \pi G \rho,
\end{equation}
where $\rho$ is the total density including all components. The radial gradient of $\psi$ is related to the rotation curve by 
\begin{equation}
\frac{v_\phi^2}{r} = \frac{\partial \psi}{\partial r},
\end{equation}
and using this in the Poisson equation we obtain
\begin{equation}
\frac{\partial g_z}{\partial z} = 4 \pi G \rho - 2 \beta \Omega^2,
\end{equation}
where $g_z = \partial \psi/\partial z$ and $\beta = d\ln v_\phi/ d\ln r$ is the rotation curve index. Integrating, we therefore have
\begin{equation}
g_z \approx \int_0^z \left(4 \pi G \rho - 2\beta \Omega^2\right) \,dz'.
\end{equation}
Note that, although it is tempting to approximate that $\beta\Omega^2$ is constant for small $z$, this approximation clearly fails for the common case of a flat rotation curve, $\beta = 0$, because $\beta = 0$ at the midplane but not above it -- see Appendix C of \citet{mckee15a} for discussion. The weight is therefore
\begin{equation}
\label{eq:wgt_integral}
\int_0^\infty \left\langle \rho_{\rm g} g_z\right\rangle \, dz = 
2\pi G \int_0^\infty \rho_{\rm g} \left[\Sigma(z) - \frac{1}{\pi G}\int_0^z \beta \Omega^2 \, dz' \right] \, dz
\end{equation}
where $\Sigma(z) = 2\int_0^z \rho \, dz$ is the total column density of material at heights between $-z$ and $z$, and we assume symmetry about $z=0$. If we write out the total column as the sum of the gas, stellar, and dark components, $\Sigma(z) = \Sigma_{\rm g}(z) + \Sigma_*(z) + \Sigma_{\rm d}(z)$, then we can integrate the gaseous part by the usual change of variables $d\Sigma_{\rm g} = 2 \rho_{\rm g} \, dz$, yielding
\begin{eqnarray}
\lefteqn{\int_0^\infty \left\langle \rho_{\rm g} g_z\right\rangle \, dz = \frac{\pi}{2} G \Sigma_{\rm g}^2 + 4\pi G}
\nonumber \\
& & {} \cdot
 \int_0^\infty \rho_{\rm g} \left(\Sigma_*(z) + \Sigma_{\rm d}(z) - \frac{1}{2\pi G} \int_0^z \beta\Omega^2\, dz'\right) \, dz.
\label{eq:wgt_integral2}
\end{eqnarray}
The dark matter scale height is much larger than the gas scale height, so we can approximate $\Sigma_d(z)  = 2 \rho_{\rm d} z$ in \autoref{eq:wgt_integral2}, where $\rho_d$ is the dark matter density inside the plane. Similarly, the stellar scale height is at least as large as the gas scale height. We can therefore use the approximation suggested by \citet{ostriker10a},
\begin{eqnarray}
\lefteqn{\int_0^\infty \left\langle \rho_{\rm g} g_z\right\rangle \, dz \approx \frac{\pi}{2} G \Sigma_{\rm g}^2 \cdot {}}
\nonumber \\
& & 
\left[1 + \frac{\zeta_{\rm d} \rho_{\rm d} + \zeta_* \rho_*}{\rhogmp}
- \frac{4}{\pi G \Sigma_{\rm g}^2} \int_0^\infty \rho_{\rm g} \int_0^z \beta\Omega^2\, dz' \, dz
\right],
\label{eq:wgt_integral3}
\end{eqnarray}
where $\rhogmp$ is the midplane gas density, $\rho_{*,\rm mp}$ is the midplane stellar density, and $\zeta_{\rm d}$ and $\zeta_*$ are numerical factors of order unity that depend on the gas density distribution and the relative scale heights of gas and stars.\footnote{Note that \citet{ostriker11a}'s equation 2 is a special case of \autoref{eq:wgt_integral3}; one can derive their equation by adopting $\beta=1$, and assuming that the angular velocity $\Omega$ arises purely from a spherical matter distribution. Also note that our $\zeta_{\rm d}$ and $\zeta_*$ differ from theirs by a factor of 4. We choose our normalisation so that $\zeta \rightarrow 1$ exactly in the limiting case where the gas and stars have the same vertical distribution} For the dark matter, which has a scale height much larger than the gas scale height, $\zeta_{\rm d}\approx 1.33$. The stellar scale height can range from much larger than that of the gas, in which case $\zeta_* \approx 1.33$ as for the dark matter, to comparable to the gas, in which case $\zeta_* \approx 1$, with exact equality holding in the case where the gas and stars have identical vertical distributions.

We therefore define
\begin{equation}
\label{eq:fgP}
f_{g,P} \equiv \left[1 + \frac{\zeta_{\rm d} \rho_{\rm d} + \zeta_* \rho_*}{\rhogmp}
- \frac{4}{\pi G \Sigma_{\rm g}^2} \int_0^\infty \rho_{\rm g} \int_0^z \beta\Omega^2\, dz' \, dz
\right]^{-1}
\end{equation}
so that
\begin{equation}
\label{eq:wgt_fgP}
\int_0^\infty \left\langle \rho_{\rm g} g_z\right\rangle \, dz = \frac{\pi}{2} G f_{g,P}^{-1}\Sigma_{\rm g}^2.
\end{equation}
The physical meaning of $f_{g,P}$ is that it is the fraction of the midplane pressure due to the local self-gravity of the gas (the unity term in \autoref{eq:fgP}), as opposed to local dark matter (as represented by the $\rho_{\rm d}$ term), local stars (as represented by the $\rho_*$ term), or material of any type interior to the radius under consideration (as represented by the $\beta\Omega^2$ term). In the Solar neighbourhood, \citet{mckee15a} obtain estimates $\rhogmp = 0.041$ $M_\odot$ pc$^{-3}$, $\rho_* = 0.043$ $M_\odot$ pc$^{-3}$, and $\rho_{\rm d} \ll \rho_*$. Using their equation 94, and adopting $\beta=0$ at the midplane, gives $1/(\pi G) \int_0^z \beta \Omega^2 \, dz' = 0.01$ $M_\odot$ pc$^{-2}$ at $z=150$ pc, approximately the gas scale height. Using these values in \autoref{eq:fgP}, and adopting $\zeta_* = 1.33$ since the stellar scale height is much larger than the gas scale height, gives $f_{g,P} \approx 0.4$ for the Solar neighbourhood, similar to $f_{g,Q}$.

Finally, inserting \autoref{eq:phimp} and \autoref{eq:wgt_fgP} into \autoref{eq:vert_hydrostatic1} gives
\begin{equation}
\rhogmp = \frac{\pi}{2 \phimp f_{g,P}} G \left(\frac{\Sigma_{\rm g}}{\sigma_{\rm g}}\right)^2.
\end{equation}
Rewriting in terms of $Q$, we arrive at our final expression for the midplane density,
\begin{equation}
\label{eq:rhomp}
\rhogmp = \frac{(1+\beta) f_{g,Q}^2}{\pi Q^2 \phimp f_{g,P}} \left(\frac{\Omega^2}{G}\right).
\end{equation}

\subsection{Energy Equilibrium}

The third assumption of our model is that gas discs are in energy equilibrium, meaning that the rate at which energy is lost due to dissipation of turbulence (ultimately leading to radiative losses) balances the rate at which it is added due to star formation feedback and input of gravitational energy due to non-axisymmetric torques. We must therefore calculate each of these three rates.

\subsubsection{Turbulent Dissipation}

Dissipation of supersonic turbulence has been subject to extensive study \citep{stone98a, mac-low98a, mac-low99b, lemaster09a}, and the consensus of this work is that the energy is lost to shocks (and, in weakly-ionised plasmas, ion-neutral friction -- \citealt{burkhart15a}) in roughly a flow crossing time at the outer scale of the turbulence. Thus the dissipation rate per unit area should be the kinetic energy per unit area divided by the crossing time. To determine the crossing time, we approximate that the outer scale of the turbulence is of order the gas scale height, and following \citet{forbes12a} we approximate this as
\begin{equation}
H_{\rm g} \approx \frac{\sigma_{\rm g}^2}{\pi G [\Sigma_{\rm g} + (\sigma_{\rm g}/\sigma_*)\Sigma_*]},
\end{equation}
where the factor $\sigma_{\rm g}/\sigma_*$ in the denominator has been chosen to interpolate between the two extreme cases where $\sigma_{\rm g}/\sigma_* \ll 1$ and $\sigma_{\rm g}/\sigma_* = 1$. In the former case,  the gas is so much thinner than the stars that the stellar distribution contributes negligibly to the vertical gravity of the gas, while in the latter case the two components have approximately the same vertical distribution. With this approximation, we can write the loss rate as
\begin{eqnarray}
\label{eq:ldiss}
\mathcal{L} & = & \eta\frac{\Sigma_{\rm g} (\sigma_{\rm g}^2-\sigma_{\rm th}^2)}{H_{\rm g}/\sqrt{\sigma_{\rm g}^2-\sigma_{\rm th}^2}}  \\
& = & \frac{2(1+\beta)}{\pi G  Q^2} \eta \phi_Q \phi_{\rm nt}^{3/2} f_{g,Q}^2 \Omega^2 \sigma_{\rm g}^3,
\label{eq:loss_rate}
\end{eqnarray}
In \autoref{eq:ldiss}, the numerator is the kinetic energy per unit area, the denominator is the scale height crossing time, and $\sigma_{\rm th}$ is the purely thermal portion of the gas velocity dispersion, which is not subject to radiative loss because the gas temperature is assumed to be set by radiative equilibrium. The quantity $\eta$ is a factor of order unity that defines the exact loss rate, with $\eta = 3/2$ corresponding to all the energy being radiated in a single scale height-crossing time; we adopt this as our fiducial value. The factors
\begin{equation}
\label{eq:phiQ}
\phi_Q \equiv 1 + \frac{Q_{\rm g}}{Q_*}
\end{equation}
and
\begin{equation}
\label{eq:phith}
\phi_{\rm nt} \equiv 1 - \frac{\sigma_{\rm th}^2}{\sigma_{\rm g}^2}
\end{equation}
are both close to unity for most galaxies. We have $\phi_Q = 2$ if $Q_{\rm g} \approx Q_*$, and we adopt this as a fiducial value. Values of $\phi_Q$ significantly greater than unity are possible only if $Q_* < Q_{\rm g}$. Similarly, the quantity $\phint$ deviates significantly from unity only for gas velocity dispersions so small that they approach the thermal velocity dispersion, which is $\approx 5$ km s$^{-1}$ in H~\textsc{i}-dominated galaxies, and $\approx 0.2 - 0.5$ km s$^{-1}$ in H$_2$-dominated ones. For most purposes we will use $\phint = 1$ as a fiducial value, corresponding to $\sigma_{\rm g} \gg \sigma_{\rm th}$, but where necessary we will evaluate $\phint$ numerically.

\subsubsection{Driving by Star Formation}
\label{sssec:driving}

Following a number of authors \citep{matzner02a, krumholz06d, krumholz17a, goldbaum11a, faucher-giguere13a}, we approximate that the rate at which star formation adds energy to the gas is determined by the asymptotic momentum of shells of gas driven by supernovae or other forms of stellar feedback. Specifically, if an energetic feedback event (such as a supernova) occurs, it will sweep up a bubble of interstellar gas that will, after all the thermal energy injected by the event has been radiated, contain  asymptotic radial momentum $p$. We approximate that this event adds an amount of energy $\approx p \sigma_{\rm g}$ to the gas when the shell breaks up and merges with the turbulence. Thus if the star formation rate per unit area is $\dot{\Sigma}_*$, and the mean momentum injected per unit mass of stars formed is $\langle p_*/m_*\rangle$, the rate of energy gain per unit area from star formation is
\begin{equation}
\calG = \left\langle\frac{p_*}{m_*}\right\rangle \sigma_{\rm g} \dot{\Sigma}_*.
\end{equation}
As discussed above, for single supernovae $\langle p_*/m_*\rangle \approx 3000$ km s$^{-1}$ \citep{cioffi88a, thornton98a, martizzi15a, kim15a, walch15b}. The momentum injected may be somewhat enhanced by clustering, though probably by at most a factor of $\sim 4$ when averaging over a realistic cluster mass function \citep{sharma14a, gentry17a, gentry18a, kim17a}. For simplicity we will ignore this effect and adopt the single supernova value $\langle p_*/m_*\rangle \approx 3000$ km s$^{-1}$ as our fiducial choice.

It is convenient to express the rate of star formation as
\begin{equation}
\label{eq:sfr}
\dot{\Sigma}_* = \epsilon_{\rm ff} f_{\rm sf} \frac{\Sigma_{\rm g}}{t_{\rm ff}}.
\end{equation}
Here $f_{\rm sf}$ is the fraction of the gas that is in a star-forming molecular phase rather than a warm atomic phase, and $t_{\rm ff}$ and $\epsilon_{\rm ff}$ are the free-fall time and star formation rate per free-fall time in this gas. As noted above, there is extensive observational evidence that $\epsilon_{\rm ff} \approx 0.01$ over a very wide range of star-forming environments \citep{krumholz07e, krumholz12a, garcia-burillo12a, evans14a, salim15a, usero15a, heyer16a, vutisalchavakul16a, leroy17a, onus18a}. We adopt $\epsilon_{\rm ff} = 0.015$, the best fit from \citet{krumholz12a}, as our fiducial choice.

We pause here to note that, in contrast to the other studies cited, \citet{lee16a}, building on the work of \citet{murray11b}, report the existence of a population of clouds with very high star formation efficiencies, $\epsilon_{\rm ff} \approx 1$. If this result were correct, it would have profound implications for models such as the one we propose. However, it is hard to reconcile this observation with the results of the numerous other studies cited above, which have failed to detect the purported high efficiency cloud population. We argue that the likely explanation for this discrepancy is a methodological bias. \citet{lee16a} compute their efficiencies based on the ratio of ionising luminosity to instantaneous gas mass. The difficulty with this technique is that the ionising luminosity is a measure of stars formed $\sim 3-5$ Myr ago, rather than the instantaneous rate at which the gas that is currently present is forming stars. The high efficiency regions that \citet{lee16a} identify are those associated with the largest and most luminous H~\textsc{ii} regions in the Milky Way, all of which have substantially disrupted their environments. \citeauthor{lee16a}'s method assumes that it is possible to map these giant bubbles one-to-one onto still-extant molecular clouds, neglecting the possibility that their present masses are not reflective of the mass of gas that went into making the ionising stars. Such a discrepancy in mass could occur because the parent clouds have been disrupted into multiple pieces by stellar feedback, or because there have been substantial flows of mass in (ongoing accretion) or out (mass loss via feedback -- \citealt{feldmann11a}) of the star forming region.

In contrast, no studies that measure star formation rates using indicators other than ionising luminosity, or that target embedded sources for which the cloud identification is much less uncertain, find a population of high efficiency clouds. Indeed, even using ionising luminosity as a star formation tracer, but in external galaxies where there is no line of sight confusion and thus it is not necessary to try to assign individual H~\textsc{ii} regions to individual molecular clouds, \citet{leroy17a} find $\epsilon_{\rm ff} \lesssim 1\%$, and with a much smaller dispersion than \citet{lee16a}. This finding strongly supports the hypothesis that \citeauthor{lee16a}'s cloud matching procedure is the source of the discrepancy between their results and the rest of the literature. For this reason, we use the value of $\epsilon_{\rm ff}$ found by all other techniques.

There is some subtlety in choosing $f_{\rm sf}$ and $t_{\rm ff}$. Some authors have simply set $f_{\rm sf} \approx 1$ and evaluated $t_{\rm ff}$ using the midplane density, and this approach is reasonable for starburst galaxies where the entire ISM is continuous, molecular, star-forming medium. However, such an approach is clearly not reasonable for galaxies like the Milky Way, where the mean density at the midplane is $\approx 1$ cm$^{-3}$, but star formation occurs exclusively in molecular clouds that constitute only $f_{\rm sf}\approx 30\%$ of the mass, but are a factor of $\gtrsim 100$ denser, giving $t_{\rm ff} \approx t_{\rm ff,mp}/10$. Indeed, such an assumption is even problematic for galaxies on the star forming main sequence at $z\sim 2$, since for some of these galaxies the midplane density implied by \autoref{eq:rhomp} is $\lesssim 10$ cm$^{-3}$. This clearly cannot all be star-forming molecular material.

In our model we follow the approach set out in \citet{forbes14a}, who base their model on the observations compiled by \citet{krumholz12a}. In this model, stars are assumed to form in a continuous medium with a free-fall time determined from $\rhogmp$ as long as the resulting star formation timescale, 
\begin{equation}
\label{eq:tsf_Toomre}
t_{\rm sf,T} \equiv \frac{t_{\rm ff}}{\epsilon_{\rm ff}} = \frac{\pi Q}{4 f_{g,Q}\epsilon_{\rm ff}}\sqrt{\frac{3 f_{g,P} \phimp}{2(1+\beta)}} \frac{1}{\Omega},
\end{equation}
is shorter than $t_{\rm sf,max} \approx 2$ Gyr, the value that appears to result in galaxies like the Milky Way where the gas breaks up into individual molecular clouds whose densities are decoupled from the mean midplane density \citep{bigiel08a, leroy08a, leroy13a}. Following the terminology of \citet{krumholz12a}, we refer to the former case as the ``Toomre regime" and the associated timescale $t_{\rm sf,T}$ defined by \autoref{eq:tsf_Toomre} as the Toomre star formation timescale, since when it applies the density in star-forming regions is set by Toomre stability of the entire disc. We refer to the latter case as the ``GMC regime", since it applies when star-forming regions have densities determined by local considerations rather than global disc stability. Thus, we take the star formation rate to be
\begin{equation}
\label{eq:sfr1}
\dot{\Sigma}_* = f_{\rm sf} \Sigma_{\rm g} \max\left(t_{\rm sf,T}^{-1}, t_{\rm sf,max}^{-1}\right),
\end{equation}
where the first case is the Toomre regime and the second is the GMC regime. In terms of the galactic orbital period, the condition for being in the Toomre regime is 
\begin{eqnarray}
\label{eq:ttoomre}
t_{\rm orb} < t_{\rm orb,T} & \equiv & \frac{8\epsilon_{\rm ff} f_{g,Q}}{Q} \sqrt{\frac{2(1+\beta)}{3 f_{g,P} \phimp}} t_{\rm sf,max} \\
& = & 35 f_{g,Q,0.5} f_{g,P,0.5}^{-1/2}\mbox{ Myr},
\end{eqnarray}
where $t_{\rm orb} = 2\pi/\Omega$ is the galactic orbital period, $f_{g,Q,0.5} \equiv f_{g,Q}/0.5$ and similarly for $f_{g,P,0.5}$, and the numerical evaluation uses the fiducial values given in \autoref{tab:quantities}. 

The value of $f_{\rm sf}$ can be computed from theoretical models \citep[e.g.,][]{krumholz09a, krumholz09b, mckee10a, krumholz13c}. For galaxies in the Toomre regime, one usually has $f_{\rm sf}\approx 1$, but this is not true for galaxies in the GMC regime. For now we choose to leave $f_{\rm sf}$ as a free parameter. Finally, using \autoref{eq:Q}, we have
\begin{eqnarray}
\mathcal{G} & = & f_{\rm sf} \frac{\sqrt{2(1+\beta)}}{\pi G Q} f_{g,Q}  \left\langle\frac{p_*}{m_*}\right\rangle \Omega \sigma_{\rm g}^2
\nonumber \\
& & \qquad {} \times \max\left(
\sqrt{\frac{32(1+\beta)}{3\pi^2 f_{g,P}}}\frac{f_{g,Q}}{Q \phimp}\epsilon_{\rm ff} \Omega,
\frac{1}{t_{\rm sf,max}}\right).
\end{eqnarray}
Note that we are implicitly neglecting other possible energy injection mechanisms, such as magnetorotational or thermal instability.

\subsection{Radial Transport}

\subsubsection{The Transport Rate Equation}

In a standard ``top down" derivation of the star formation law, the next step would be to equate the rates of loss from turbulent dissipation $\calL$ and gain from star formation feedback $\calG$. Since these have different scalings -- $\calL \propto \Omega^2 \sigma_{\rm g}^3$ and $\calG \propto \epsilon_{\rm ff} \Omega^2 \sigma_{\rm g}^2$ (in the Toomre regime) or $\calG \propto t_{\rm sf,max}^{-1} \Omega \sigma_{\rm g}^2$ (in the GMC regime) -- such equality can hold everywhere within the disc only if $\sigma_{\rm g}$ takes on a particular, fixed value (and hence $Q$ is non-constant), or if $\epsilon_{\rm ff}$ is non-constant. For example, \citet{ostriker11a} make the former choice, while \citet{faucher-giguere13a} and \citet{hayward17a} make the latter. Neither option provides a particularly good match to observations, for the reasons discussed in \autoref{sec:intro}.

Our model is based on the realisation that there is an alternative source of energy, radial transport. Such transport injects energy at scales comparable to the gas scale height, which then cascades down to become turbulent on smaller scales. \citet{krumholz10c} show that the time evolution of the gas velocity dispersion obeys
\begin{eqnarray}
\frac{\partial\sigma_{\rm g}}{\partial t} & = & \frac{\calG-\calL}{3\sigma_{\rm g}\Sigma_{\rm g}} + \frac{\sigma_{\rm g}}{6\pi r\Sigma_{\rm g}} \frac{\partial \dot{M}}{\partial r}
+ \frac{5(\partial\sigma_{\rm g}/\partial r)}{6\pi r\Sigma}\dot{M}
\nonumber \\
& & \qquad {} - \frac{1-\beta}{6\pi r^2\Sigma_{\rm g}\sigma_{\rm g}}\Omega \calT,
\label{eq:encons}
\end{eqnarray}
where $\calT$ is the torque exerted by non-axisymmetric stresses, and
\begin{equation}
\label{eq:mdot}
\dot{M}= -\frac{1}{v_\phi(1+\beta)}\frac{\partial \calT}{\partial r}
\end{equation}
is the rate of inward mass accretion through the disc. Note that $\dot{M}$ here and throughout refers explicitly to mass accretion \textit{through} the disc rather than \textit{onto} the disc from outside, unless explicitly stated otherwise. There is clear physical interpretation for  \autoref{eq:encons}. The first term on the right hand side is the net effect of star formation driving ($\mathcal{G}$) and dissipation of turbulence ($\mathcal{L}$), the second and third represent advection of kinetic energy as gas moves through the disc, and the final term represents transfer of energy from the galactic gravitational potential to the gas.

We pause here to comment on the physical assumptions that lie behind \autoref{eq:encons}. This equation is simply the time- and azimuthally-averaged version of the equation of energy conservation for a thin disc with a time-steady rotation curve, and it holds regardless of the nature of the torque $\mathcal{T}$. Thus it can apply equally well to gas transport driven by transient or steady spiral waves (as in a modern galaxy) or transport coming from the mutual torquing of giant clumps (as in a high-$z$ galaxy). However, \autoref{eq:encons} does not include another energy source that is at least in principle possible: transfer of energy from stars to gas without any transport of the gas itself, for example due to stellar spiral arms or bars directly driving turbulent gas motions. That is, it is possible to ``pay" for an increase in gas kinetic energy by having the stars decrease their energy by either flowing down the potential well or decreasing their velocity dispersion, and such transfer could take place even if gas does not flow down the potential well, or even flows up it. Such star-to-gas direct transfer probably is important in some regions, particularly those with little gas and strong bars, as discussed in \autoref{ssec:limitations}. However, numerous numerical simulations of both local \citep{agertz09b, goldbaum15a, goldbaum16a} and high-$z$ \citep{bournaud07a, bournaud09a, ceverino10a} galaxies offer strong evidence that direct star-to-gas energy transfer cannot be a dominant source of gas kinetic energy. These simulations show that gas does flow inward at roughly the rate predicted by our model, even when a live stellar disc and its spiral waves are included in the simulations, and, conversely, that turbulence and inflow occur even in simulations that do not include a massive stellar disc. Neither of these findings is consistent with the hypothesis that stellar driving rather than gas transport dominates the energy budget in most galaxies.

If we search for solutions where that gas is in energy equilibrium, $\partial\sigma_{\rm g}/\partial t = 0$, then \autoref{eq:encons} implies that
\begin{equation}
\label{eq:eneq}
\frac{\sigma_{\rm g}^2}{2\pi r} \frac{\partial \dot{M}}{\partial r} + \frac{5 \sigma_{\rm g} \dot{M}}{2\pi r}  \frac{\partial \sigma_{\rm g}}{\partial r} - \frac{1-\beta}{2\pi r^2} \Omega \calT = \calL - \calG
\end{equation}
This is a second order ordinary differential equation in $\calT$ (since $d\dot{M}/dr$ involves the second derivative of $\calT$), with $\calL - \calG$ as a forcing term. Physically valid solutions to this equation are subject to the constraint $\calT \rightarrow 0$ as $r\rightarrow 0$, so that no torques are exerted (and thus no energy is added) at $r=0$.

\subsubsection{The Critical Velocity Dispersion}

Following \citet{forbes14a}, we note that the solutions to this equation are only consistent with thermodynamic constraints when $\calL \geq \calG$, i.e., when the dissipation of turbulence is stronger than driving, so the forcing term is positive. If this inequality holds, then gravitationally-driven turbulence transports mass inward and converts gravitational potential energy into turbulent motion at the rate required to maintain the gas in a state of marginal stability. In the opposite case, however, gravitational instability would be required to convert energy from random motions into a net outward transport of mass, which is unphysical on thermodynamic grounds -- the turbulence is assumed to be randomly oriented, so there is plausible physical mechanism by which it could self-organise to generate a net outward mass transport. If $\calL = \calG$ exactly, then driving by star formation is by itself sufficient to offset the decay of turbulence, and there is no gravitational instability or radial transport.

The condition that $\calL = \calG$ for a marginally stable disc with $Q = \Qmin$ is satisfied if the gas velocity dispersion (total thermal plus non-thermal) is
\begin{eqnarray}
\label{eq:sigma_sf}
\sigma_{\rm g} = \sigma_{\rm sf} & \equiv &
\frac{4f_{\rm sf} \epsilon_{\rm ff}}{\sqrt{3 f_{g,P}} \pi \eta \phimp \phi_Q \phint^{3/2}}
\left\langle\frac{p_*}{m_*}\right\rangle
\nonumber \\
& & 
\;{} \times
\max\left[1,
\sqrt{\frac{3 f_{g,P}}{8(1+\beta)}} \frac{\Qmin \phimp}{4 f_{g,Q} \epsilon_{\rm ff}}\frac{t_{\rm orb}}{t_{\rm sf,max}}\right].
\end{eqnarray}
With this definition, we can rewrite the equation for energy equilibrium, \autoref{eq:eneq}, as
\begin{equation}
\label{eq:eneq1}
\frac{\sigma_{\rm g}^2}{2\pi r} \frac{\partial \dot{M}}{\partial r} + \frac{5 \sigma_{\rm g} \dot{M}}{2\pi r}  \frac{\partial \sigma_{\rm g}}{\partial r} - \frac{1-\beta}{2\pi r^2} \Omega \calT = \calL \left(1-\frac{\sigma_{\rm sf}}{\sigma_{\rm g}}\right).
\end{equation}
With the energy equation written in this way, the physical meaning of $\sigma_{\rm sf}$ becomes clear. It is the velocity dispersion that star formation alone is capable of maintaining, without any additional energy input from mass transport. As the velocity dispersion of the ISM approaches this limit, the net rate of turbulent dissipation diminishes, and the amount of gravitational transport required to maintain marginal stability does as well. The fraction of the energy supplied by star formation is simply $\sigma_{\rm sf}/\sigma_{\rm g}$, while the fraction supplied by gravity is $1-\sigma_{\rm sf}/\sigma_{\rm g}$. Once the galaxy reaches $\sigma_{\rm g} = \sigma_{\rm sf}$ exactly, the mass inflow rate drops to 0, and the galaxy is no longer constrained to have $Q = \Qmin$; it can instead take on any value of $Q \geq \Qmin$.

We can also express the condition that $\calL = \calG$, and thus gravitational power shut off, in terms of the surface density. Combining \autoref{eq:Qdef} and \autoref{eq:sigma_sf}, the critical surface density at which this occurs is
\begin{eqnarray}
\label{eq:Sigma_sf}
\Sigma_{\rm sf}  & = & 
\frac{8\sqrt{2(1+\beta)} f_{\rm sf} \epsilon_{\rm ff}}{\sqrt{3} \pi G Q \eta \phi_{\rm mp} \phi_Q \phi_{\rm nt}^{3/2}} \left\langle\frac{p_*}{m_*}\right\rangle \frac{f_{g,Q}}{f_{g,P}^{1/2}} \frac{1}{t_{\rm orb}}
\nonumber \\
& &
\quad \cdot {} \max\left[1, \sqrt{\frac{3 f_{g,P}}{2(1+\beta)}} \frac{Q \phi_{\rm mp}}{8 f_{g,Q} \epsilon_{\rm ff}} \frac{t_{\rm orb}}{t_{\rm sf,max}}\right].
\end{eqnarray}
Transport shuts off wherever $\Sigma_{\rm g}$ falls below $\Sigma_{\rm sf}$. Note that higher values of $Q$ imply lower values of $\Sigma_{\rm sf}$, i.e., the more gravitationally stable the disc, the lower the total surface density that can be maintained by star formation alone. The maximum surface density that can be sustained by star formation alone in a marginally stable disc is given by $\Sigma_{\rm sf}$ evaluated with $Q = \Qmin$.

Numerical evaluation of \autoref{eq:sigma_sf} and \autoref{eq:Sigma_sf} requires some care due to the $\phi_{\rm nt}$ term in the denominator. Our fiducial choice for this term is $\phi_{\rm nt} = 1$, appropriate for highly-supersonic gas ($\sigma_{\rm sf} \gg \sigma_{\rm th}$). In most cases this choice is not problematic. However, one regime of interest for our theory is  H~\textsc{i}-dominated regions like the outer Milky Way or the majority of $z=0$ dwarfs, which have $f_{\rm sf} \ll 1$. Examination of \autoref{eq:sigma_sf} would seem to suggest that sufficiently small values of $f_{\rm sf}$ will produce correspondingly small values of $\sigma_{\rm sf}$, in which case the approximation that $\sigma_{\rm sf} \gg \sigma_{\rm th}$, and thus $\phi_{\rm nt} \approx 1$, is no longer valid; indeed, for $\sigma_{\rm sf} \rightarrow \sigma_{\rm th}$ we have $\phi_{\rm nt} \rightarrow 0$. Thus we cannot simply assume $\phi_{\rm nt} = 1$ when evaluating \autoref{eq:sigma_sf} for H~\textsc{i}-dominated regions; a more sophisticated approach is required.

If one substitutes the full definition $\phi_{\rm nt} = 1 - (\sigma_{\rm th} / \sigma_{\rm g})^2$ into \autoref{eq:sigma_sf}, the resulting equation is a cubic in $\sigma_{\rm g}^2$. While we can solve this exactly, the solution is extremely cumbersome and unenlightening. It is more useful to obtain the solution in the two limiting cases $\sigma_{\rm g} \gg \sigma_{\rm th}$ and $\sigma_{\rm g} \rightarrow \sigma_{\rm th}$; numerical solution of the full cubic shows that $\sigma_{\rm sf}$ transitions smoothly between the two limits. The solution for $\sigma_{\rm g} \gg \sigma_{\rm th}$ is simply what we would have obtained by naively plugging in $\phi_{\rm nt} = 1$, which is
\begin{eqnarray}
\sigma_{\rm sf} & = & 11\mbox{ km s}^{-1} f_{\rm sf} f_{g,P,0.5}^{-1/2}
\nonumber \\
& &
\quad {} \cdot \max\left(1, 1.0 f_{g,P,0.5}^{1/2} f_{g,Q,0.5}^{-1} t_{\rm orb,100}\right),
\label{eq:sigma_sf_num1}
\\
\Sigma_{\rm sf} & = & 36 \, M_\odot\mbox{ pc}^{-2} f_{\rm sf} f_{g,Q,0.5} f_{g,P,0.5}^{-1/2} t_{\rm orb,100}^{-1} 
\nonumber \\
& &
\quad {} \cdot \max\left(1, 1.0 f_{g,P,0.5}^{1/2} f_{g,Q,0.5}^{-1} t_{\rm orb,100}\right),
\end{eqnarray}
where $t_{\rm orb,100} = t_{\rm orb}/100$ Myr, for $\Sigma_{\rm sf}$ we have used $Q=\Qmin$, and for all quantities we have used the fiducial parameter choices as given in \autoref{tab:quantities}.

We treat the $\sigma_{\rm sf} \approx \sigma_{\rm th}$ limit by defining the Mach number $\mathcal{M}_{\rm sf}$ corresponding to $\sigma_{\rm sf}$ by
\begin{equation}
\sigma_{\rm sf} = \sigma_{\rm th}\sqrt{1 +\mathcal{M}_{\rm sf}^2}
\end{equation}
(so that $\sigma_{\rm th} = \sigma_{\rm sf}$ corresponds to $\mathcal{M}_{\rm sf} = 0$), and solve \autoref{eq:sigma_sf} to first order in $\mathcal{M}_{\rm sf}$. This gives
\begin{eqnarray}
\mathcal{M}_{\rm sf} & = & \left\{\frac{4f_{\rm sf} \epsilon_{\rm ff}}{\sqrt{3 f_{g,P}} \pi \eta \phimp \phi_Q}
\left\langle\frac{p_*}{m_*}\right\rangle\frac{1}{\sigma_{\rm th}}
\right.
\nonumber \\
& & 
\quad \left.
{} \cdot
\max\left[1,
\sqrt{\frac{3 f_{g,P}}{8(1+\beta)}} \frac{\Qmin \phimp}{4 f_{g,Q} \epsilon_{\rm ff}}\frac{t_{\rm orb}}{t_{\rm sf,max}}\right]\right\}^{1/3} \\
& = & 0.60 f_{\rm sf,0.1}^{1/3} f_{g,P,0.5}^{1/6} \sigma_{\rm th,5}^{-1/3}
\nonumber \\
& &
\quad {} \cdot \max\left(1, 1.0 f_{g,P,0.5}^{1/6} f_{g,Q,0.5}^{-1/3} t_{\rm orb,100}^{1/3}\right),
\label{eq:sigma_sf_num2}
\\
\Sigma_{\rm sf} & = & \frac{\sqrt{8(1+\beta)} f_{g,Q} \sigma_{\rm th}}{G \Qmin t_{\rm orb}} \\
& = & 16\,M_\odot\mbox{ pc}^{-2} f_{g,Q,0.5} \; \sigma_{\rm th,5} t_{\rm orb,100}^{-1}
\end{eqnarray}
where $\sigma_{\rm th,5} = \sigma_{\rm th}/5$ km s$^{-1}$. Thus we find that, for the relatively modest star-forming fractions typical of H~\textsc{i}-dominated regions, the maximum Mach number that can be sustained by star-formation is of order $0.5$. Since $\sigma_{\rm th} \approx 5$ km s$^{-1}$ in the warm neutral medium, this in turn implies overall velocity dispersions of $\approx 6-8$ km s$^{-1}$. Thus we find that, regardless of the value of $f_{\rm sf}$ or various other parameters, our model predicts that the maximum velocity dispersion that can be sustained by star formation alone is $\sigma_{\rm sf} \approx 6-10$ km s$^{-1}$. A corollary of this statement is that, if we observe a galaxy's velocity dispersion to be close to $\sigma_{\rm sf}$, we can conclude that the turbulence within it is primarily powered by star formation, whereas if we observe the velocity dispersion to be $\gg \sigma_{\rm sf}$, we can conclude that the turbulence is primarily powered by gravity. We also note that our finding that star formation at a rate consistent with the observed Kennicutt-Schmidt relation is capable of powering a velocity dispersion of $\approx 10$ km s$^{-1}$ and no more is not new; several numerical simulations of supernova-driven turbulence have reached the same conclusion from their numerical experiments \citep[e.g.,][]{joung09a, kim11a, kim15b}.

\subsubsection{The Steady-State Mass Inflow Rate}
\label{sssec:mdot_ss}

With $\sigma_{\rm sf}$ defined, we are now in a position to calculate the mass inflow rate for galaxies with $\sigma_{\rm g} > \sigma_{\rm sf}$ and $Q = \Qmin$. \citet{krumholz10c} obtained a transport equation analogous to \autoref{eq:eneq1} in the limit $\sigma_{\rm g} \gg \sigma_{\rm sf}$, and for constant $\beta$ (i.e., fixed rotation curve index) showed that it admits an analytic steady state solution with $\sigma_{\rm g}$ and $\dot{M}$ independent of radius. Numerical solution of the full time-dependent system (\autoref{eq:encons}) shows that galaxies tend to approach this steady state \citep{forbes12a, forbes14a}, so motivated by this result we look for similar solutions ($\beta$, $\sigma$, $\dot{M}$ all independent of $r$) for the more general case given by \autoref{eq:eneq1}.\footnote{An important subtlety: in writing \autoref{eq:sigma_sf_num2} we evaluated $\phi_{\rm nt}$ using $\sigma_{\rm g} = \sigma_{\rm sf}$. This is the correct approach to finding the value of $\sigma_{\rm sf}$ that can be sustained by star formation alone. However, in \autoref{eq:eneq1}, $\sigma_{\rm sf}$ must be evaluated using the actual value of $\sigma_{\rm g}$, which may be larger.} A solution of this form must have $\calT = -\dot{M} v_\phi r$, and inserting this into \autoref{eq:eneq1} we immediately obtain that the mass inflow rate must be
\begin{eqnarray}
\label{eq:mdot_steady}
\dot{M} = \dot{M}_{\rm ss} & \equiv & \frac{4 (1+\beta) \eta \phi_Q \phint^{3/2}}{(1-\beta) G \Qmin^2} f_{g,Q}^2 \sigma_{\rm g}^3 \left(1 - \frac{\sigma_{\rm sf}}{\sigma_{\rm g}}\right) \\
& = & 0.71 f_{g,Q,0.5}^2 \sigma_{\rm g,10}^3\;M_\odot\mbox{ yr}^{-1}
\nonumber \\
& & {} \cdot
 \left(1 - \frac{\sigma_{\rm th}^2}{\sigma_{\rm g}^2}\right)^{3/2} \left(1 - \frac{\sigma_{\rm sf}}{\sigma_{\rm g}}\right),
\end{eqnarray}
where $\sigma_{\rm g,10} = \sigma_{\rm g}/10$ km s$^{-1}$; the numerical evaluation uses the fiducial values in \autoref{tab:quantities}, except that we have retained the explicit dependence on $\phi_{\rm nt}$ because it is important in H~\textsc{i}-dominated regions, as explained above. The quantity $\dot{M}_{\rm ss}$ is the steady-state mass inflow rate that is required to keep a galactic disc in energy equilibrium. Thus we expect galactic discs with $\sigma_{\rm g} \approx 10$ km s$^{-1}$, and thus slightly above $\sigma_{\rm sf}$ and $\sigma_{\rm th}$ to have mass inflow rates of order $1$ $M_\odot$ yr$^{-1}$. As $\sigma_{\rm g}$ decreases and approaches both $\sigma_{\rm sf}$ and $\sigma_{\rm th}$, the inflow rate rapidly falls to zero, while as it increases the inflow rate rises as $\dot{M}_{\rm ss} \propto \sigma_{\rm g}^3$.

\begin{figure*}
\includegraphics[width=0.8\textwidth]{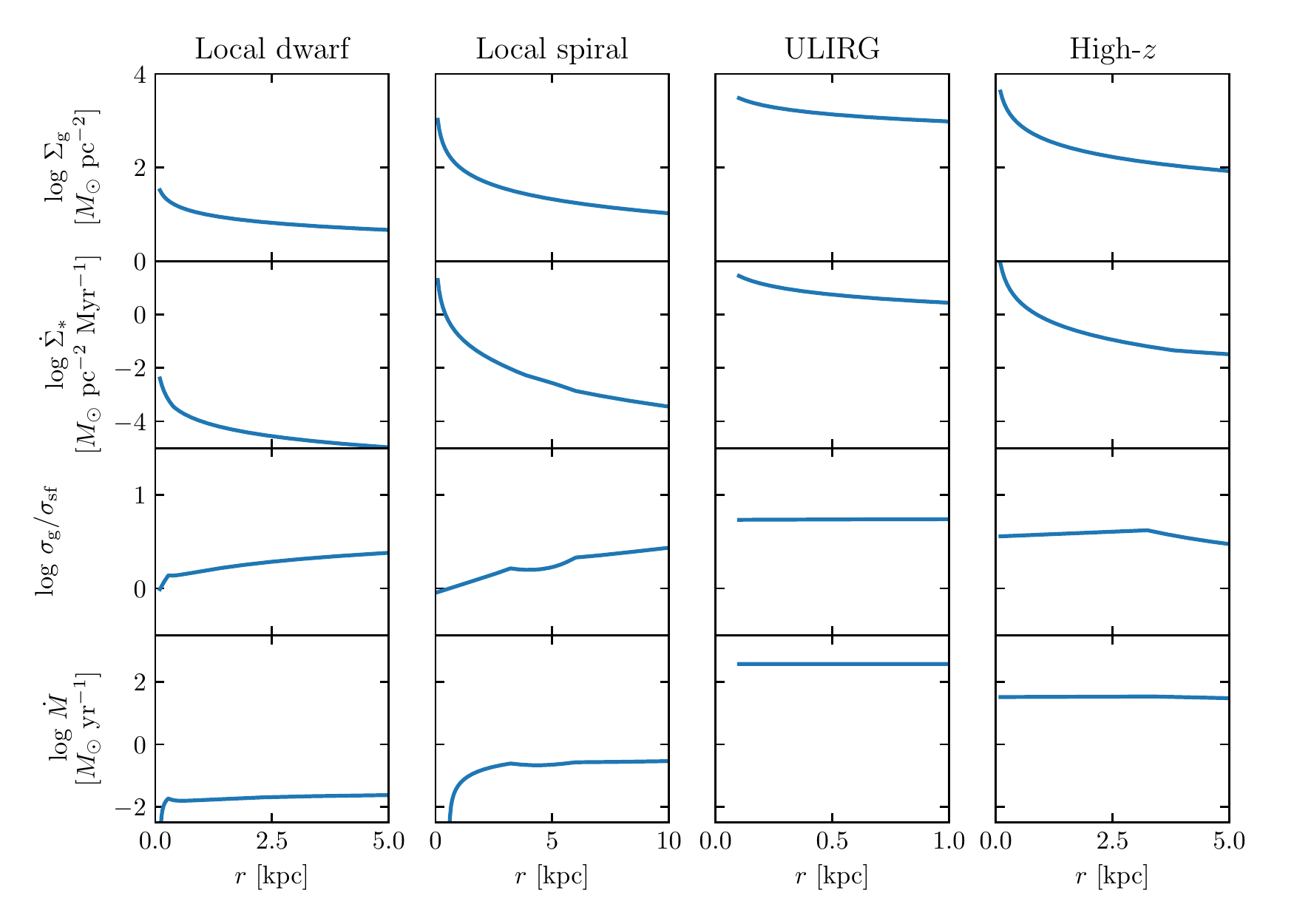}
\caption{
\label{fig:example_sol}
Example solutions for our fiducial model, using the parameters chosen for local dwarfs, local spirals, ULIRGs, and high-redshift star-forming galaxies given in \autoref{tab:example_sol}. Note that different columns have different $x$ axis ranges. Rows show, from top to bottom, gas surface density $\Sigma_{\rm g}$, star formation surface density $\dot{\Sigma}_*$, ratio of gas velocity dispersion $\sigma_{\rm g}$ to dispersion provided by star formation $\sigma_{\rm sf}$, and mass inflow rate $\dot{M}$.
}
\end{figure*}

\begin{table}
\begin{tabular}{c@{$\quad$}cccc}
\hline
Parameter & Local dwarf & Local spiral & ULIRG & High-$z$ \\
\hline
$\sigma_{\rm g}$ [km s$^{-1}$] & 6 & 10 & 60 & 40 \\
$r_{\rm out}$ [kpc] & 5 & 10 & 1 & 5 \\
$v_{\phi}$ at $r_{\rm out}$ [km s$^{-1}$] & 60 & 200 & 250 & 200 \\
$\beta$ & 0.5 & 0 & 0.5 & 0 \\
$Z'$ & 0.2 & 1 & 1 & 1 \\
\hline
\end{tabular}
\caption{
\label{tab:example_sol}
Parameters for example solutions. Note that $r_{\rm out}$ is the outermost radius at which we compute the solution, and $Z'$ is the metallicity normalised to Solar used in the KMT+ model for $f_{\rm sf}$ (see main text).
}
\end{table}

We show some example equilibrium solutions in \autoref{fig:example_sol}; the examples are representative of the range of galaxies to which we can apply our model, including a local dwarf, a local spiral similar to the Milky Way, a local ULIRG, and a high redshift star-forming disc. The exact parameters for each model are given in \autoref{tab:example_sol}. All models use $f_{g,Q} = f_{g,P} = 0.5$, and an inner radius of $0.1$ kpc. We use a value of $f_{\rm sf}$ computed using the KMT+ model of \citet{krumholz13c} with a clumping factor $f_c = 1$, since the gas surface densities here are the true ones rather than a beam-diluted average. To apply this theory we require a value for the midplane stellar plus dark matter density. If the rotation curve index is independent of radius, and is dominated by stars and dark matter, then the minimum density at the midplane required to produce the rotation curve is
\begin{equation}
\label{eq:rho_min}
\rho_{\mathrm{min}} = \frac{v_\phi^2}{4\pi G r^2} \left(2\beta+1\right).
\end{equation}
The true value is likely to be somewhat higher, since $\rho_{\rm min}$ applies for a spherical mass distribution, which we would expect if dark matter alone were dominating the rotation curve; we therefore adopt a stellar density $\rho_{\rm *} = 2 \rho_{\rm min}$. We compute the thermal velocity dispersion $\sigma_{\rm th}$ as $\sigma_{\rm th} = f_{\rm sf} \sigma_{\rm th,mol} + (1-f_{\rm sf})\sigma_{\rm th, WNM}$, where $\sigma_{\rm th,mol} = 0.2$ km s$^{-1}$ (appropriate for molecular gas) and $\sigma_{\rm th,WNM} = 5.4$ km s$^{-1}$ appropriate for warm neutral gas. The results illustrate the qualitative behaviour of the model: local spirals and dwarfs with modest velocity dispersions and modest star formation rates have $\sigma_{\rm g}/\sigma_{\rm sf} \approx 1$, and as a result also have low mass inflow rates, $\sim 10^{-2}$ $M_\odot$ yr$^{-1}$ for the dwarf and $\sim 1$ $M_\odot$ yr$^{-1}$ for the spiral. In contrast, rapidly star-forming ULIRGs and high-redshift galaxies have high $\sigma_{\rm g}/\sigma_{\rm sf}$ and high inflow rates. The turbulence in these galaxies is driven almost entirely by inflow.

\subsection{Equilibria without Transport or without Feedback}

It is worth considering the alternatives to our model that result from omitting either feedback or transport, in order to demonstrate why both are important. First consider omitting feedback, as in \citet{krumholz10c}. This amounts to setting $\langle p_*/m_*\rangle = 0$, and thus all the relations we have derived continue to apply, but with $\sigma_{\rm sf} = 0$ and $\Sigma_{\rm sf} = 0$. 

The other alternative is models without transport, which require that $\calG = \calL$. As noted above, this requirement can be satisfied in two ways. One is that we can keep the star formation law (\autoref{eq:sfr}) fixed. In the GMC regime we have $\calG \propto \Sigma_{\rm g} \sigma_{\rm g}$ while $\calL \propto \Sigma_{\rm g}^2 \sigma_{\rm g}$, and thus $\calG = \calL$ is possible only for a single value of $\Sigma_{\rm g}$; since real galaxies clearly do not all have a single surface density, we discount this solution and instead focus on the Toomre regime. In the Toomre regime we have $\calG = \calL$ whenever $\sigma_{\rm g} = \sigma_{\rm sf}$ (\autoref{eq:sigma_sf}). This implies that
\begin{eqnarray}
Q & = & f_{g,Q} \frac{\kappa \sigma_{\rm sf}}{\pi G \Sigma_{\rm g}} 
\nonumber \\
\label{eq:Q_notransport}
& = & \frac{8\sqrt{2(1+\beta)}}{\sqrt{3}\pi \eta \phimp \phi_Q \phint^{3/2}} f_{\rm sf} \epsilon_{\rm ff} \left\langle\frac{p_*}{m_*}\right\rangle \frac{f_{g,Q}}{f_{g,P}^{1/2} G \Sigma_{\rm g} t_{\rm orb}} \\
& = & 3.6 f_{g,Q,0.5} f_{g,P,0.5}^{-1/2} t_{\rm orb,100}^{-1} \Sigma_{\rm g,10}^{-1},
\end{eqnarray}
where $\Sigma_{\rm g,10} = \Sigma_{\rm g}/10$ $M_\odot$ pc$^{-2}$. Thus if we do not include transport and keep the star formation law fixed, the model still predicts that $Q\approx 1$ for Solar Circle conditions ($\Sigma_{\rm g,10} \approx 1$, $t_{\rm orb,100} \approx 2$). However, for conditions like those found in ULIRGs ($\Sigma_{\rm g,10}\sim 100$, $t_{\rm orb,10} \sim 3$) or high-$z$ star-forming discs ($\Sigma_{\rm g,10} \sim 10$, $t_{\rm orb,100} \sim 1$), the predicted value of $Q$ is much smaller than unity.

Conversely, we can hold $Q$ fixed and treat the quantity $f_{\rm sf} \epsilon_{\rm ff}$ as a free parameter, and use the relation $\calG = \calL$ to solve for it. In this case only the Toomre regime exists, and it is characterised by a star formation efficiency per free-fall time
\begin{eqnarray}
\label{eq:epsff_fg}
\epsilon_{\rm ff} & = & \frac{\sqrt{3} \pi \eta \phimp \phi_Q \phint^{3/2}}{4 f_{\rm sf}} \left\langle\frac{p_*}{m_*}\right\rangle^{-1} f_{g,P}^{-1/2} \sigma_{\rm g} \\
& = & 0.027 f_{\rm sf}^{-1} f_{g,P,0.5}^{-1/2} \sigma_{\rm g,10}
\end{eqnarray}
Thus $\epsilon_{\rm ff}$ is $\sim 1\%$ for $\sigma_{\rm g} \approx 10$ km s$^{-1}$, but rises to $\gtrsim 10\%$ for the higher velocity dispersions typically seen in ULIRGs or high-redshift star-forming discs. Note that \autoref{eq:epsff_fg} is identical, up to factors of order unity, to equation 37 of \citet{faucher-giguere13a}.

\section{Comparison to Observations}
\label{sec:observations}

We can use our steady state model to calculate a wide range of observables, and in this section we compare the model predictions to observations. We also compare contrasting models without transport and without feedback, in order to highlight how including both mechanisms alters the results. Specifically, throughout this section we will consider four different models, to which we refer as follows:

\textbf{Transport+feedback.} This is our fiducial model. It has $\epsilon_{\rm ff} = 0.015$ and two branches: $Q = \Qmin$ with $\sigma_{\rm g} > \sigma_{\rm sf}$ (or equivalently $\Sigma_{\rm g} > \Sigma_{\rm sf}$), and $Q \geq \Qmin$ with $\sigma_{\rm g} = \sigma_{\rm sf}$ (or $\Sigma_{\rm g} \leq \Sigma_{\rm sf}$). 

\textbf{No-feedback.} This is identical to the transport+feedback model, except that $\sigma_{\rm sf} = 0$ and $\Sigma_{\rm sf} = 0$, so $Q = \Qmin$ under all circumstances. This model is similar to the one proposed by \citet{krumholz10c}.

\textbf{No-transport, fixed $\epsilon_{\rm ff}$.} A model without transport, with $\epsilon_{\rm ff} = 0.015$ fixed but $Q$ allowed to vary freely. In this model the value of $Q$ is given by \autoref{eq:Q_notransport}. This model is similar to the one proposed by \citet{ostriker11a}.

\textbf{No-transport, fixed $Q$.} A model without transport, with $Q = \Qmin$ fixed, but $\epsilon_{\rm ff}$ allowed to vary freely. In this model, $\epsilon_{\rm ff}$ takes on the value given by \autoref{eq:epsff_fg}. This model is similar to the one proposed by \citet{faucher-giguere13a}.

For each of these models we compute the star-forming fraction $f_{\rm sf}$ using the formalism of \citet{krumholz13c}, with a clumping factor $f_c = 5$ (since we are now dealing with beam-diluted kpc-scale observations), Solar metallicity, and a stellar density equal to 4 times the minimum value given in \autoref{eq:rho_min}.

\subsection{The Star Formation Law}
\label{ssec:sflaw}

A first test of any model of star formation is the prediction it makes for the star formation law, the relation between the gas content of galaxies and their star formation behaviour. Observationally, the star formation law can be expressed as a correlation between the surface density of star formation and either the gas surface density alone, or the gas surface density divided by the galactic orbital period. It can be measured averaged over entire galaxies, or measured in spatially-resolved patches of galaxies. A successful model should be able to reproduce all these observed correlations.\footnote{One can also define a local star formation law, which relates the local rate of star formation within a given cloud to its volumetric properties (density, virial ratio, etc.). There are significant observational constraints on this relationship as well, as discussed in \autoref{sssec:driving}, but in this paper we have used these constraints as an input to the model, not an output, and thus our model cannot be said to predict this relation. However, the local volumetric star formation relation is distinct from the projected, area-averaged one, and it is perfectly possible to match observations of one without successfully reproducing the other. Indeed, in the following sections we will encounter a number of models that do exactly that. Thus the models we consider do constitute predictions for the areal star formation law.}

\subsubsection{Spatially-Resolved Observations}

First consider spatially-resolved observations. For both the transport+feedback model and the no-feedback model, the star formation rate at each point in the disc is described by \autoref{eq:sfr1} with $\epsilon_{\rm ff} = 0.015$. If we omit star formation feedback, only the $Q = \Qmin$ solution branch exists, whereas in our fiducial transport+feedback model we can have $Q > \Qmin$ for $\Sigma_{\rm g} < \Sigma_{\rm sf}$. (Recall that we are limiting our attention to discs in energy equilibrium without significant external energy input; external stimulation can produce $Q \gg Q_{\rm min}$ -- \citealt{inoue16a}.) In practice, however, this makes relatively little difference in the star formation law unless we adopt $Q \gg \Qmin$, though we shall see that it makes a considerable difference for other observables. Thus for simplicity we simply adopt $Q = \Qmin$ everywhere, in which case the transport+feedback and no-feedback models are the same.

In the no-transport, fixed $\epsilon_{\rm ff}$ model, the value of $Q$ is given by \autoref{eq:Q_notransport}. Substituting this into \autoref{eq:sfr1} (and recalling that the GMC regime does not exist in this case) gives a star formation law
\begin{equation}
\label{eq:sflaw_observed_os}
\dot{\Sigma}_* = \pi G \eta \phimp^{1/2} \phi_Q \phint^{3/2} \left\langle\frac{p_*}{m_*}\right\rangle^{-1} \Sigma_{\rm g}^2.
\end{equation}
This relation is identical up to factors of order unity to equation 10 of \citet{ostriker11a}, which is not surprising since it is based on the same physical assumptions. 

In the no-transport, fixed $Q$ model, we instead have $Q=\Qmin$ and a value of $\epsilon_{\rm ff}$ given by \autoref{eq:epsff_fg}. Inserting this into \autoref{eq:sfr1} gives
\begin{equation}
\label{eq:sflaw_observed_fqh}
\dot{\Sigma}_* = \pi^2 G \eta \phimp^{1/2} \phi_Q \phint^{3/2} f_{g,P}^{-1} \Qmin \left\langle\frac{p_*}{m_*}\right\rangle^{-1} \Sigma_{\rm g}^2,
\end{equation}
with no dependence on the orbital period. This equation is identical up to factors of order unity with equation 18 of \citet{faucher-giguere13a}. It is also nearly identical to \autoref{eq:sflaw_observed_os} -- the scalings are the same, and the leading coefficients differ only by a factor of $\pi \Qmin/f_{g,P} \sim 1$.\footnote{Despite the fact that \autoref{eq:sflaw_observed_os} and \autoref{eq:sflaw_observed_fqh} make nearly identical predictions for the star formation law, the routes by which they arrive at these predictions are quite different. In deriving \autoref{eq:sflaw_observed_os}, one assumes that the star formation efficiency per free-fall time is constant. The scaling $\dot{\Sigma}_* \propto \Sigma_{\rm g}^2$, implying a star formation timescale that declines as $t_{\rm sf}\propto 1/\Sigma_{\rm g}$, arises because the gas velocity dispersion is constant, and this leads to a midplane density that increases as the square of $\Sigma_{\rm g}$. This in turn leads to a free-fall time that scales as $\Sigma_{\rm g}^{-1}$. In contrast, in deriving \autoref{eq:sflaw_observed_fqh} one assumes that the midplane density is not varying, since it is fixed by the condition $Q=\Qmin$. Instead, the efficiency of star formation is proportional to $\Sigma_{\rm g}$. Thus \autoref{eq:sflaw_observed_os} corresponds to a picture where the star formation process is not sensitive to the gas surface density in a galaxy, but the midplane density is, while \autoref{eq:sflaw_observed_fqh} arises from a picture where the midplane density is independent of gas surface density, but the star formation process is not.} Thus for the purposes of comparing to observation we need only consider one form of the no-transport model. An important point to note is that the factor $f_{\rm sf}$ vanishes in both \autoref{eq:sflaw_observed_os} and \autoref{eq:sflaw_observed_fqh}, as it must, since in these models the star formation rate always self-adjusts to maintain force and energy balance without any help from transport. 

\begin{figure}
\includegraphics[width=\columnwidth]{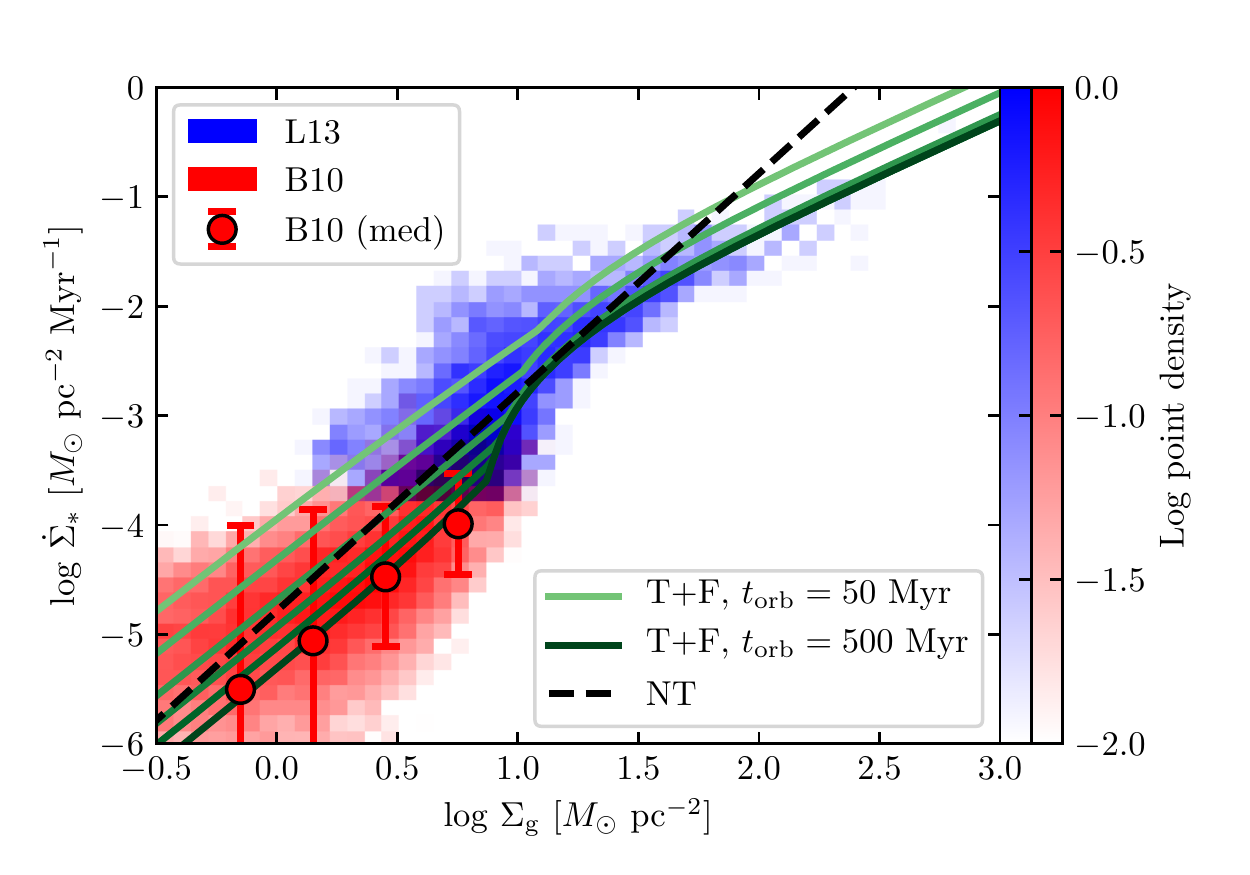}
\caption{
\label{fig:sflaw_resolved}
Comparison between theoretical models predictions of the star formation law and observation of nearby galaxies at $\sim 1$ kpc resolution. Lines represent models; solid green lines are the transport+feedback model (\autoref{eq:sfr1}; T+F in the legend), evaluated for orbital times evenly spaced in logarithm from $t_{\rm orb} = 50 - 500$ Myr, with lighter colours (toward the top) corresponding to shorter orbital times. The dashed black line is the no-transport model (\autoref{eq:sflaw_observed_os}; NT in the legend), which has no dependence on orbital time. All models use the fiducial parameters given in \autoref{tab:quantities}, and we compute the star-forming molecular fraction $f_{\rm sf}$ from the KMT+ model  \citep{krumholz13c} as in \autoref{sssec:mdot_ss}, using a Solar-normalised metallicity $Z' = 1/3$, appropriate for dwarfs and outer discs. Coloured histograms show observations; colours indicate the distribution of individual pixels in the $\Sigma_{\rm g} - \dot{\Sigma}_*$ plane for inner galaxies (blue; \citealt{leroy13a}; L10 in the legend) and outer galaxies and dwarfs (red; \citealt{bigiel10a}; B10 in the legend); red circles with error bars show the median and scatter of the outer galaxy data. 
}
\end{figure}

We therefore have two prospective predictions of the star formation law to consider: our fiducial transport+feedback model (\autoref{eq:sfr1} evaluated with $Q=\Qmin$), and a no-transport model  (\autoref{eq:sflaw_observed_os}). We plot the model predictions together with resolved observations in \autoref{fig:sflaw_resolved}. The fiducial model does a good job of describing the data for plausible input values of $t_{\rm orb}$ -- the range plotted is $50-500$ Myr, which roughly covers the span of the data, which include regions from galactic centres to outskirts. In particular, the fiducial model properly captures the curvature seen in the data, where the slope of $\dot{\Sigma}_*$ versus $\Sigma_{\rm g}$ is clearly steeper in the range $\log(\Sigma_{\rm g}/M_\odot\,\mathrm{pc}^{-2}) \approx 0.5 - 1$ than at either higher or lower surface density. In comparison, the no-transport model produces noticeably too steep a slope compared to the observations. The mismatch is most apparent at surface densities of $\sim 100$ $M_\odot$ pc$^{-2}$, where a model without transport tends to over-predict the star formation rate by more than an order of magnitude. Moreover, the no-transport model is unable to reproduce the curvature of the data associated with the atomic- to molecular-dominated transition at $\approx 10$ $M_\odot$ pc$^{-2}$, because the star formation rate is insensitive to the thermal or chemical state of the ISM in this case.

\subsubsection{Unresolved Observations}
\label{sssec:sflaw_unresolved}

\begin{figure*}
\includegraphics[width=0.8\textwidth]{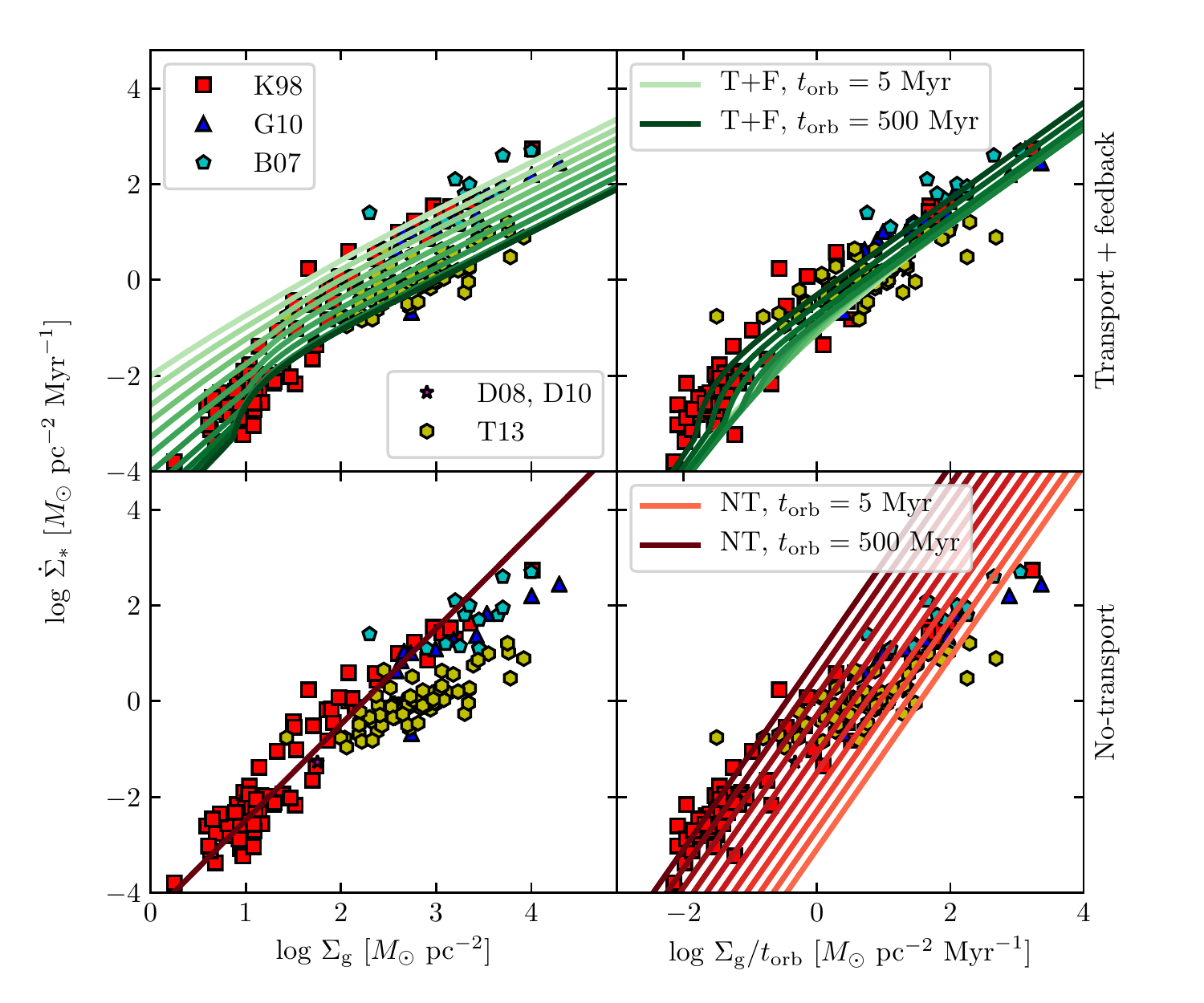}
\caption{
\label{fig:sflaw_unresolved}
Comparison between theoretical model predictions of the star formation law and observations of marginally-resolved galaxies, with one measurement per galaxy. In each panel, the $y$ axis shows the star formation rate per unit area $\dot{\Sigma}_*$. In the left column the $x$ axis shows total gas surface density $\Sigma$, while in the right it shows $\Sigma/t_{\rm orb}$. The top row shows our fiducial transport+feedback (T+F) model (\autoref{eq:sfr_averaged}), while the bottom row shows the no-transport (NT) model (\autoref{eq:sflaw_observed_os}). Coloured lines show the model predictions, evaluated using orbital periods from $t_{\rm orb} = 5$ Myr (lighter colours) to $500$ Myr (darker colours), with lines evenly spaced in $\log t_{\rm orb}$; note that only one line appears in the lower left panel, because the relationship between $\Sigma_{\rm g}$ and $\dot{\Sigma}_*$ is independent of $t_{\rm orb}$ in the no-transport model. All plots have star-forming ISM fraction $f_{\rm sf}$ computed from the KMT+ model as in \autoref{ssec:sflaw}, and use the fiducial values given in \autoref{tab:quantities}, except that we use $\beta=0.5$ rather than 0 because a substantial part of the sample consists of circumuclear starbursts which are in regions with rising rather than flat rotation curves. Coloured points, which are the same in each panel, show data culled from the following sources: local galaxies from \citet[][K98 in the legend]{kennicutt98a}, $z\sim 2$ sub-mm galaxies from \citet{bouche07a}[][B07 in the legend], and galaxies on and somewhat above the star-forming main sequence at $z\sim 1 - 3$ from \citet[][D08, D10 in the legend]{daddi08a, daddi10a}, \citet[][G10 in the legend]{genzel10a}, and \citet[][T13 in the legend]{tacconi13a}. The observations have been homogenised to a \citet{chabrier05a} IMF and the convention for $\alpha_{\rm CO}$ suggested by \citet{daddi10b}; see \autoref{app:alpha_CO} and \citet{krumholz12a} for details.
}
\end{figure*}

For unresolved observations, we have access only to the surface densities of gas and star formation averaged over the entire disc, and to the rotation period at the disc edge. To compare our model to such data, we must take care to average the model predictions in the same way. Doing so precisely requires knowing the radial variation of the gas surface density and all the other factors in \autoref{eq:sfr1}, which is obviously not possible for unresolved observations. However, we can make a rough estimate for the effects of area-averaging by considering a disc with radially-constant values of the gas velocity dispersion $\sigma_{\rm g}$, rotation curve index $\beta$, stability parameter $Q$, the various gas fractions $f_{g,Q}$ and $f_{g,P}$, and the star-forming fraction $f_{\rm sf}$. From \autoref{eq:Qdef}, we can see that such a a disc has a surface density that varies with radius as $\Sigma_{\rm g} \propto v_\phi/r \propto r^{\beta-1}$. Thus if the disc extends from inner radius 0 to some finite outer edge, the area-averaged surface density is larger than the surface density at the edge by a factor of $2/(1+\beta)$. 

The effects of area-averaging on the star formation rate depend on the star formation law. First consider our transport+feedback case or a case with no feedback, both of which follow \autoref{eq:sfr1}. In discs where the majority of the star formation occurs in the GMC regime, where the star formation timescale is constant, the area-averaged star formation rate is larger than the value at the outer edge by the same factor. However, in portions of the disc in the Toomre regime, \autoref{eq:sfr1} gives a star formation rate per unit area that varies as $\dot{\Sigma}_* \propto \Sigma_{\rm g} \Omega \propto r^{2(\beta-1)}$. For $\beta \neq 1$, this gives an area-averaged star formation surface density that is larger than the value at the disc edge by a factor of $1/\beta$. Thus the area-averaged version of \autoref{eq:sfr1} can be written
\begin{equation}
\label{eq:sfr_averaged}
\langle\dot{\Sigma}_*\rangle \approx f_{\rm sf} \langle \Sigma_{\rm g}\rangle \phi_a \max\left[\frac{4 \epsilon_{\rm ff} f_{g,Q}}{\pi Q} \sqrt{\frac{2(1+\beta)}{3 f_{g,P} \phimp}} \Omega_{\rm out}, t_{\rm sf,max}^{-1}\right],
\end{equation}
where the angle brackets indicate area averages, and $\Omega_{\rm out}$ is the angular velocity at the outer edge of the star-forming disc.

The factor $\phi_a$ represents the difference in the factors by which area-averaging enhances the star formation rate compared to the gas surface density. It is unity for discs in the GMC regime; in the Toomre regime it is $(1+\beta)/2\beta$ for $\beta \neq 0$. The case of a flat rotation curve, $\beta = 0$, requires special consideration, since in the Toomre regime such a disc has a total star formation rate that diverges logarithmically near the disc centre. As noted by \citet{krumholz16a}, this divergence is a result of the unphysical assumption that a flat rotation curve can continue all the way to $r=0$; such a rotation curve has a divergent shear, which in turn makes the midplane density required to maintain constant $Q$, and thus the total star formation rate, diverge. If one instead considers the more realistic case of a rotation curve that is flat only to some finite inner radius $r_0$, then the area-averaged star formation rate is larger than the value at the disc edge at radius $r_1$ by a factor of $2\ln (r_1/r_0)$, and thus $\phi_a = \ln (r_1/r_0)$. In practice this factor cannot be that large, because extended discs with flat rotation curves also tend to have much of their star formation in the GMC regime, where this extra enhancement does not occur. For this reason, we will adopt $\phi_a = 2$ as a fiducial value, recognising that it can be somewhat larger or smaller depending on the rotation curve and how much of the disc is in the Toomre regime.

We can proceed analogously to derive the offsets between the local and disc-averaged star formation laws for the alternative no-transport models. In the no-transport, fixed $Q$ model, the star formation law obeys $\dot{\Sigma}_* \propto \Sigma_{\rm g}^2$ (\autoref{eq:sflaw_observed_fqh}), we again have $\dot{\Sigma}_* \propto r^{2(\beta-1)}$, and the factor $\phi_a$ is therefore the same as in the transport+feedback case. In the no-transport, fixed $\epsilon_{\rm ff}$ model (\autoref{eq:sflaw_observed_os}), we cannot calculate the run of $\Sigma_{\rm g}$ versus radius from our assumptions, because the values of gas surface denstiy $\Sigma_{\rm g}$ and velocity dispersion $\sigma_{\rm g}$ are independent of one another. Thus we cannot directly calculate $\phi_a$ without making an additional assumption about the radial variation of $\Sigma_{\rm g}$. For simplicity, however, we will assume the same radial variation as in the $Q=\Qmin$ models, and thus obtain the same $\phi_a$. Thus the area-averaged versions of \autoref{eq:sflaw_observed_os} or \autoref{eq:sflaw_observed_fqh} are identical to the original versions, with an added factor of $\phi_a$ on the right hand side.

We compare the model predictions to a sample of unresolved observations culled from the literature in  \autoref{fig:sflaw_unresolved}. In plotting the data we use the CO-H$_2$ conversion factor $\alpha_{\rm CO}$ recommended by \citet{daddi10a}, and we discuss this choice further in \autoref{app:alpha_CO}. We see that the transport+feedback model agrees reasonably well with the data, while the no-transport model produces noticeably too steep a slope in both $\dot{\Sigma}_*$ versus $\Sigma_{\rm g}$ and $\dot{\Sigma}_*$ versus $\Sigma_{\rm g}/t_{\rm orb}$. The model including transport fares significantly better.

\subsection{Gas Velocity Dispersions}
\label{ssec:sfr_vdisp}

A second observable that we can predict is the gas velocity dispersions in galaxies, and its correlation with star formation. Consider a galaxy with a constant gas velocity dispersion $\sigma_{\rm g}$. Using the star formation relation \autoref{eq:sfr1} and our definition of $Q$ (\autoref{eq:Qdef}), we can write the star formation rate per unit area as
\begin{eqnarray}
\dot{\Sigma}_* & = & f_{\rm sf} \frac{\sqrt{8(1+\beta)} f_{g,Q}}{G Q} \frac{\sigma_{\rm g}}{t_{\rm orb}^2}
\nonumber \\
& & 
\quad {} \cdot
\max \left[\frac{8\epsilon_{\rm ff} f_{g,Q}}{Q} \sqrt{\frac{2(1+\beta)}{3f_{g,P} \phi_{\rm mp}}}, \frac{t_{\rm orb}}{t_{\rm sf,max}}\right].
\label{eq:sigma_sfr_obs}
\end{eqnarray}
As in \autoref{sssec:sflaw_unresolved}, we can derive an unresolved version of this relation under the assumption that $Q$, $f_{\rm sf}$, and $\beta$ are constant with radius. Integrating over radius, we find that the total star formation rate is
\begin{eqnarray}
\dot{M}_* & = & 
\sqrt{\frac{2}{1+\beta}} \frac{\phi_a f_{\rm sf}}{\pi G Q} f_{g,Q} v_{\phi,\rm out}^2 \sigma_{\rm g}
\nonumber \\
& &
\quad {} \cdot
\max\left[
\sqrt{\frac{2(1+\beta)}{3 f_{g,P} \phi_{\rm mp}}} \frac{8 \epsilon_{\rm ff} f_{g,Q}}{Q},
\frac{t_{\rm orb,out}}{t_{\rm sf,max}}
\right],
\label{eq:sigma_sfr_unresolved}
\end{eqnarray}
where $v_{\phi,\rm out}$ and $t_{\rm orb,out}$ are the circular velocity and orbital period evaluated at the outer edge of the star-forming disc. 

In our transport+feedback model, \autoref{eq:sigma_sfr_obs} and \autoref{eq:sigma_sfr_unresolved} are to be evaluated with $Q = \Qmin$ if $\sigma_{\rm g} > \sigma_{\rm sf}$. If $\sigma_{\rm g} = \sigma_{\rm sf}$, then we can have any $Q \geq Q_{\rm min}$. Finally, values of $\sigma_{\rm g} < \sigma_{\rm sf}$ are not possible in equilibrium.

Our alternative models have a variety of other behaviours. In the no-feedback model \autoref{eq:sigma_sfr_obs} and \autoref{eq:sigma_sfr_unresolved} are the same, but with $\sigma_{\rm sf} = 0$, and thus $Q = \Qmin$ for all $\sigma_{\rm g}$, and all values of $\sigma_{\rm g}$ are allowed. Conversely, in the no-transport, fixed $\epsilon_{\rm ff}$ model, $\sigma_{\rm g}$ can only take on the one value $\sigma_{\rm sf}$; no other values are allowed in equilibrium, and $\dot{\Sigma}_*$  are $\dot{M}_*$ are independent of this. Finally, in the no-transport, fixed $Q$ model, we have $Q = \Qmin$, and we must use \autoref{eq:epsff_fg} for $\epsilon_{\rm ff}$. Substituting this value of $\epsilon_{\rm ff}$ into \autoref{eq:sigma_sfr_obs} and \autoref{eq:sigma_sfr_unresolved} gives the relationships
\begin{eqnarray}
\dot{\Sigma}_* & = & \frac{8(\beta+1)\pi \eta \sqrt{\phi_{\rm mp} \phi_{\rm nt}^3} \phi_Q}{G Q^2 \langle p_*/m_*\rangle f_{g,P}} \frac{\sigma_{\rm g}^2}{t_{\rm orb}^2} \\
\dot{M}_* & = & \frac{4 \eta \sqrt{\phi_{\rm mp} \phi_{\rm nt}^3} \phi_Q \phi_a}{G Q^2 \langle p_*/m_*\rangle} \frac{f_{g,Q}^2}{f_{g,P}} v_{\phi,\rm out}^2 \sigma_{\rm g}^2.
\end{eqnarray}

Note that the transport+feedback and no-feedback models both predict $\dot{M}_* \propto \sigma_{\rm g}$ for $\sigma_{\rm g} > \sigma_{\rm sf}$, while the two no-transport models predict very different scalings: no relationship between $\dot{M}_*$ and $\sigma_{\rm g}$ for the no-transport, fixed $\epsilon_{\rm ff}$ model, and a much stronger scaling, $\dot{M}_* \propto \sigma_{\rm g}^2$, for the no-transport, fixed $Q$ model. This difference, first pointed out by \citet{krumholz16a}, provides a very clear observational signatures that can be used to distinguish models with and without transport.\footnote{The scaling between $\dot{M}_*$ and gas fraction for the no-transport, fixed $Q$ model that we obtain here is slightly different from that given in \citet{krumholz16a}, because here we have treated this model as having fixed total $Q$. In contrast, the  \citet{faucher-giguere13a} model to which \citet{krumholz16a} compare assumed fixed $Q_{\rm g}$.} The physical origin of this difference is easy to understand. The star formation rate is $\dot{\Sigma}_* = \epsilon_{\rm ff} \Sigma_{\rm g}/t_{\rm ff}$. For a fixed rotation curve, orbital time, and gas fraction, the gas surface density scales as $\Sigma_{\rm g} \propto \sigma_{\rm g} / Q$, and the midplane density scales as $\rho_{\rm mp} \propto Q^{-2}$, implying that the free-fall time scales as $t_{\rm ff} \propto Q$, with no explicit dependence on $\sigma_{\rm g}$. The overall scaling is therefore $\dot{\Sigma}_* \propto \epsilon_{\rm ff} \sigma_{\rm g} / Q^2$. The difference between the transport+feedback and no-transport models then follows from their assumed variations in $\epsilon_{\rm ff}$ and $Q$. Our fiducial transport+feedback model has $\epsilon_{\rm ff}$ and $Q$ both constant, so we obtain a linear scaling $\dot{\Sigma}_* \propto \sigma_{\rm g}$. The no-transport, fixed $\epsilon_{\rm ff}$ model has constant $\sigma_{\rm g}$ and varying $Q$, so it predicts no relationship between $\dot{\Sigma}_*$ and $\sigma_{\rm g}$, with all the variations in star formation rate being driven by changes in $Q$. The no-transport, fixed $Q$ model has $\epsilon_{\rm ff} \propto \sigma_{\rm g}$ (\autoref{eq:epsff_fg}), so it predicts $\dot{\Sigma}_* \propto \sigma_{\rm g}^2$.

\begin{table}
\begin{tabular}{c@{$\quad$}cccc}
\hline
Parameter & Local dwarf & Local spiral & ULIRG & High-$z$ \\
\hline
$f_{\rm sf}$ & 0.2 & 0.5 & 1.0 & 1.0 \\
$v_\phi$ [km s$^{-1}$] & 100 & 220 & 300 & 200 \\
$t_{\rm orb}$ [Myr] & 100 & 200 & 5 & 200 \\
$\beta$ & 0.5 & 0.0 & 0.5 & 0.0 \\
$f_{g,Q} = f_{g,P}$ & 0.9 & 0.5 & 1.0 & 0.7 \\
$\phi_a$ & 1 & 1 & 2 & 3 \\
$\dot{M}_{*,\rm min}$ [$M_\odot$ yr$^{-1}$] & - & - & 1 & 1 \\
$\dot{M}_{*,\rm max}$ [$M_\odot$ yr$^{-1}$] & 0.5 & 5 & - & - \\
\hline
\end{tabular}
\caption{
\label{tab:sfrvdisp_param}
Parameter values used for the theoretical models shown in \autoref{fig:sigma_sfr_unresolved}. See main text for details.
}
\end{table}

\begin{figure*}
\includegraphics{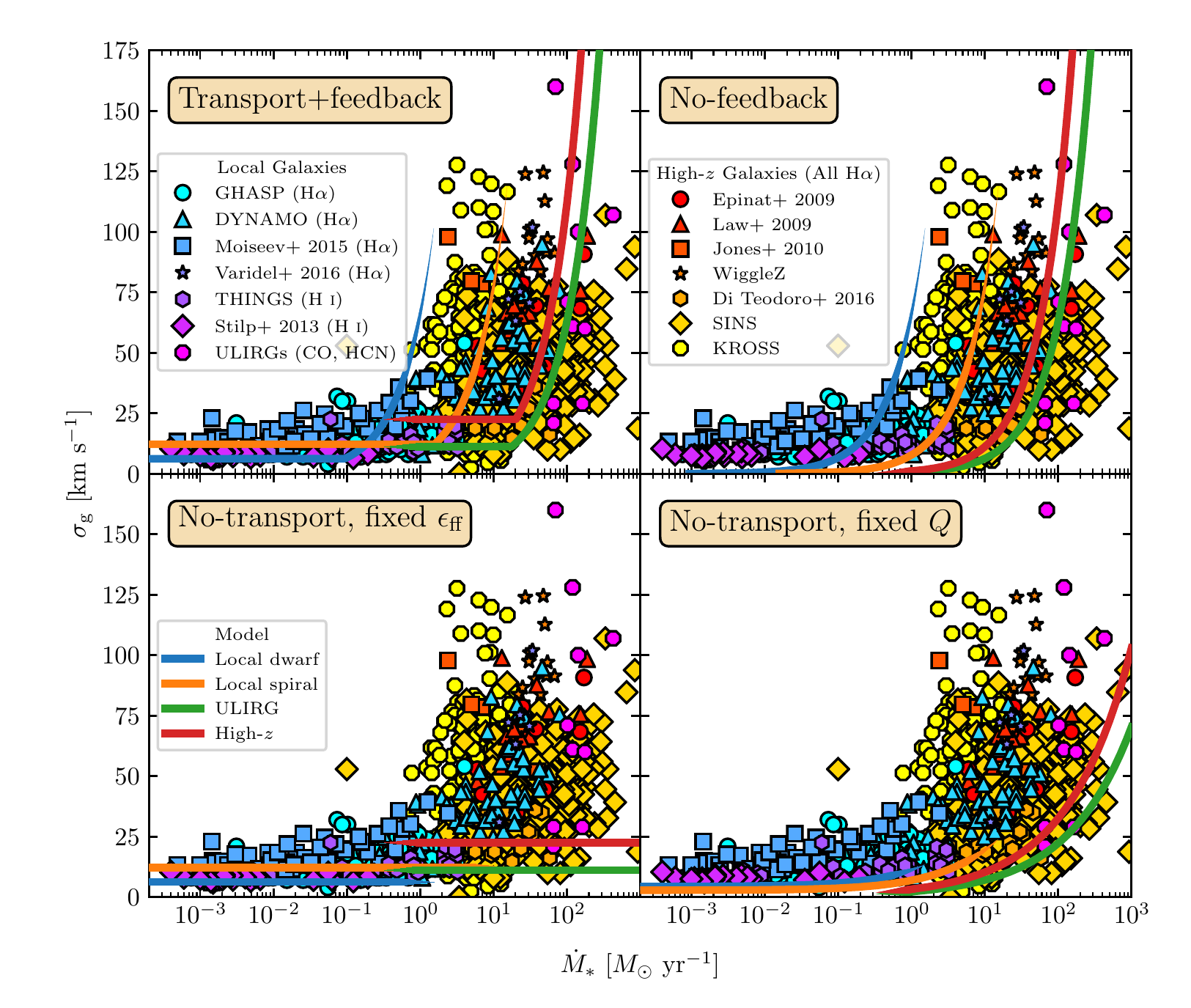}
\caption{
\label{fig:sigma_sfr_unresolved}
Comparison between the observed correlation between gas velocity dispersion and star formation rate and theoretical models. Solid lines represent theoretical models, with the model plotted indicated in each panel; clockwise from top left, these are the transport+feedback model, the no-feedback model, the no-transport, fixed $Q$ model, and the no-transport, fixed $\epsilon_{\rm ff}$ model. The lines shown are for four representative sets of parameters, corresponding roughly to those appropriate for local dwarfs galaxies, local spiral galaxies, ULIRGs, and high-$z$ star-forming discs; the lines fade outside the range of star formation rates for which they are applicable. See \autoref{tab:sfrvdisp_param} and the main text for details. The coloured points represent observations, and are the same in every panel. Data shown in include: H$\alpha$ observations of local galaxies from two surveys (GHASP, \citealt{epinat08a}, and DYNAMO, \citealt{green14a}) as well as smaller studies \citep{moiseev15a, varidel16a}; H~\textsc{i} observations of nearby galaxies from THINGS \citep{leroy08a, walter08a, ianjamasimanana12a} and from the survey of dwarfs by \citet{stilp13a}; a compilation of molecular line observations of nearby ULIRGS \citep{downes98a, sanders03a, veilleux09a, scoville15a, scoville17a}; H$\alpha$ observations of high-redshift galaxies from the samples of \citet{epinat09a}, \citet{law09a}, \citet{lemoine-busserolle10a}, and the WiggleZ \citep{wisnioski11a} and SINS-KMOS-3D \citep{wisnioski15a, wuyts16a} surveys at $z\sim 1-3$; H$\alpha$ observations of lensed galaxies at $z\sim 2-3$ from \citet{jones10a}; a sample of galaxies at $z\sim 1$ from the KMOS survey \citep{wisnioski15a} as analysed by \citet{di-teodoro16a}, and a sample from the KROSS survey \citep{stott16a, johnson17a}. Full details on the data set are given in \autoref{app:sigma_sfr_data}.
}
\end{figure*}

To compare the various theoretical models to observations, since the results depend on $f_{\rm sf}$, $v_\phi$, $t_{\rm orb}$, $\beta$, and the gas fraction, we must choose values for these parameters. For unresolved observations, the data will also depend on $\phi_a$. Following our approach in \autoref{sssec:mdot_ss}, we consider four different possibilities that should be broadly representative of the ranges these parameters can take. We label these cases local dwarf, local spiral, ULIRG, high-$z$, with the final case intended to be typical of the observed high-redshift star-forming discs. We summarise the chosen parameters in \autoref{tab:sfrvdisp_param}; all other parameters have their fiducial values as specified in \autoref{tab:quantities}, and we compute the ISM thermal velocity dispersion $\sigma_{\rm th}$ as in \autoref{sssec:mdot_ss}. Broadly speaking, the dwarf is characterised by a high gas fraction, a low star-forming fraction, and a low orbital velocity, and has $\phi_a = 1$ because it is entirely in the GMC regime; the ULIRG has a high orbital velocity, a high gas fraction, and a short orbital period. It has a larger value of $\phi_a$, since it is entirely in the Toomre regime. The local spiral and high-$z$ star-forming disc have properties intermediate between these extremes, with the high-$z$ system having a higher gas fraction, star-forming fraction, and $\phi_a$. Finally, we note that each set of model parameters is found only in some finite range of star formation rates; for example, objects with $f_{\rm sf} = 0.2$, as we adopt for our local dwarf case, do not generally produce star formation rates of $10$ $M_\odot$ yr$^{-1}$. For this reason we use each set of properties only up to some maximum or down to some minimum star formation rate. We give the limiting values $\dot{M}_{*,\rm min}$ and $\dot{M}_{*,\rm max}$ in \autoref{tab:sfrvdisp_param} as well.

We compare the predictions of our transport+feedback model and the three alternatives to observations in \autoref{fig:sigma_sfr_unresolved}. Details on the observations and our processing of them are given in \autoref{app:sigma_sfr_data}. We see that the transport+feedback model is in generally good agreement with the observations at both low and high star formation rates. In particular, it captures the behaviour that the velocity dispersion reaches a floor of $\approx 10$ km s$^{-1}$ at low star formation rates,\footnote{Some of the data, particularly the GHASP and \citet{moiseev15a} samples, have $\sigma_{\rm g} \approx 20$ km s$^{-1}$ at low star formation rates, but this is likely an artefact of using H$\alpha$-estimates of $\sigma_{\rm g}$, as the other tracers are all systematically lower. See \autoref{app:sigma_sfr_data} for further discussion.} while increasing rapidly with star formation rate at star formation rates above a few $M_\odot$ yr$^{-1}$. (The dwarf case rises to high velocity dispersions at star formation rates that are too low, but this is an artefact of choosing to fix $f_{\rm sf}$, when in fact no galaxy with a star formation rate of $\gtrsim 1$ $M_\odot$ yr$^{-1}$ has $f_{\rm sf} = 0.2$ and $v_\phi = 100$ km s$^{-1}$, as we have adopted in the dwarf case.) In contrast, the alternative models all have an obvious failing. The no-feedback model does well at high star formation rates, but fails to capture the floor imposed by star formation at low star formation rates, instead predicting that the velocity dispersion should fall to very small values. Conversely, the no-transport, fixed $\epsilon_{\rm ff}$ model correctly captures the behaviour at low star formation rates, but fails to reproduce the observed increase in velocity dispersion at higher star formation rates. Finally, the no-transport, fixed $Q$ model has qualitatively correct behaviour, but seriously under-predicts the velocity dispersion at all star formation rates. This failure is a direct result of having too steep a relationship between star formation and gas surface density, as seen in \autoref{ssec:sflaw}.

\subsection{Mass Transport}
\label{ssec:obs_transport}

A final observable, or at least potential observable, that we can predict is the correlation between mass inflow rate and physical properties of the star-forming disc. As discussed in \autoref{sec:intro}, at present we have direct detections of inflow rates only for a handful of nearby galaxies, but we can compare our model to these, and predict the results of future observations and simulations.

To make predictions for this correlation using our transport+feedback model, for any choice of $\sigma_{\rm g}$ and ancillary parameters (gas fraction, rotation curve, etc.), we can use \autoref{eq:mdot_steady} to compute the mass inflow rate, and \autoref{eq:sigma_sfr_unresolved} to compute the corresponding star formation rate. When $\sigma_{\rm g} \gg \sigma_{\rm sf}$, this leads to a predicted scaling between inflow and star formation $\dot{M} \propto \dot{M}_*^3/(v_\phi^6 f_{\rm sf}^3)$, with the coefficient depending on the gas fraction, rotation curve index, and whether the galaxy is in the Toomre regime. We can use the same method for the no-feedback model simply by setting $\sigma_{\rm sf} = 0$, but the results are only slightly different, so we refrain from showing them. Models without transport, depending on one's perspective, either predict that the inflow rate should be zero or make no predictions at all regarding its value.

\begin{figure}
\includegraphics[width=\columnwidth]{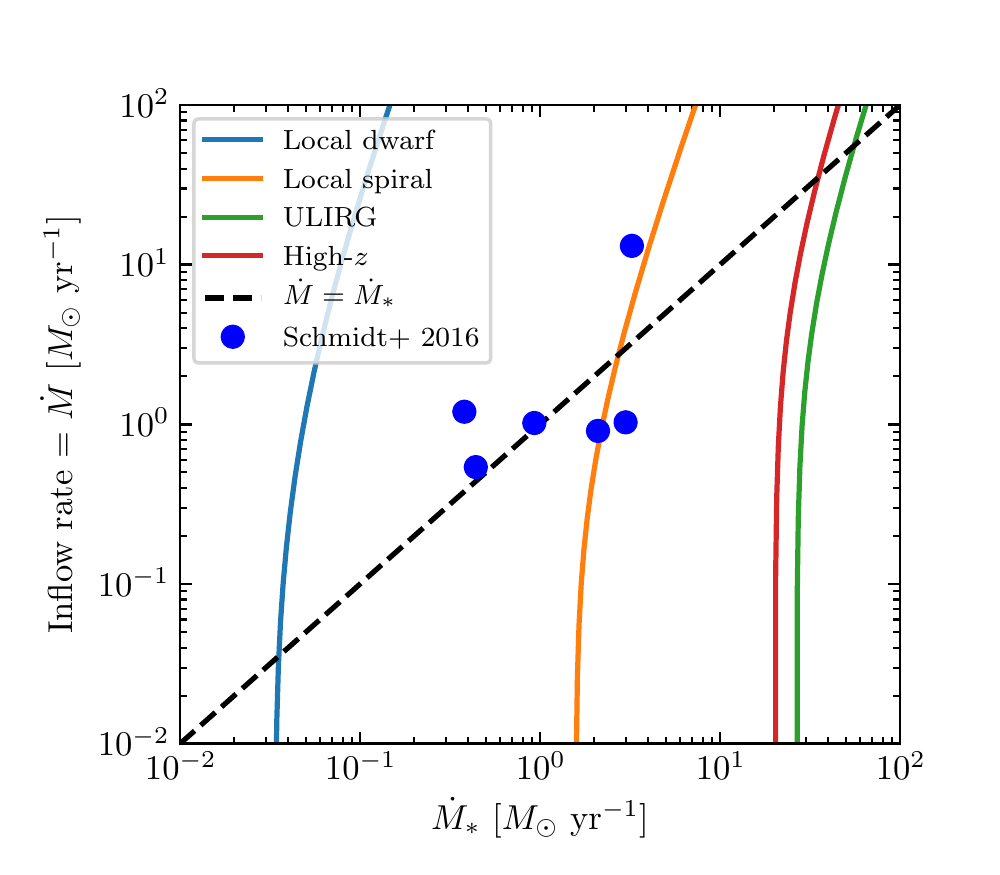}
\caption{
\label{fig:inflow_sfr}
Inflow rate through this disc versus star formation rate. Blue points show the observations of \citet{schmidt16a}, while lines show the predictions of our transport+feedback model for the four example cases whose parameters are given in \autoref{tab:sfrvdisp_param}.
}
\end{figure}

\begin{figure}
\includegraphics[width=\columnwidth]{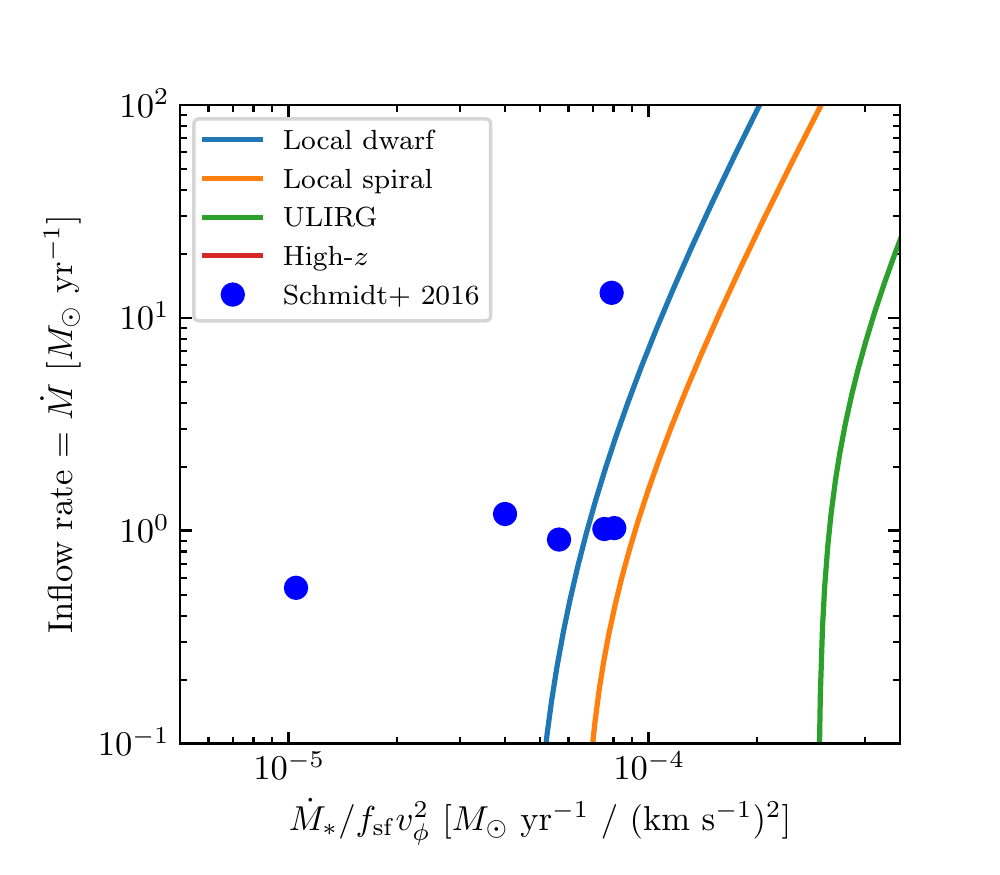}
\caption{
\label{fig:inflow_sfr_mod}
Same as \autoref{fig:inflow_sfr}, but with the star formation rate normalised by $f_{\rm sf} v_\phi^2$. Note that, compared to \autoref{fig:inflow_sfr}, the $x$ axis range has been compressed from 4 dex to 2 dex.
}
\end{figure}

We show predicted mass inflow rates for the same four example cases used in \autoref{ssec:sfr_vdisp} in \autoref{fig:inflow_sfr}. Given the extremely strong scaling of the inflow rate with $v_\phi$ and $f_{\rm sf}$, it is not surprising that the example cases cover a very wide range of possible inflow rates for a given star formation rate. Thus the model is consistent with the data, in that the data lie near the ``local spiral" parameter choices, where we expect them, but this is a relatively weak statement.

A more interesting test is to normalise out the dependence on the rotation curve velocity and star forming fraction. Our model predicts that $\dot{M}_* \propto f_{\rm sf} v_\phi^2$, whereas $\dot{M}$ is independent of these two parameters (except very close to $\sigma_{\rm g} = \sigma_{\rm sf}$), and thus our model makes a much stronger prediction for the correlation of $\dot{M}$ with $\dot{M}_*/(f_{\rm sf} v_\phi^2)$. For the observations, we take $v_{\phi}$ from Table 2 of \citet{leroy13a}; as a proxy for the star-forming fraction $f_{\rm sf}$, we use $f_{\rm sf} = \dot{\Sigma}_* / (\Sigma_{\rm g}/2\mbox{ Gyr})$, where the values of $\Sigma_{\rm g}$ and $\dot{\Sigma}_*$ are taken from the same table.

We plot the correlation of $\dot{M}$ with $\dot{M}_*/(f_{\rm sf} v_\phi^2)$ in \autoref{fig:inflow_sfr_mod}. We see that both the data and the models cluster much more tightly than in the plot of $\dot{M}$ versus $\dot{M}_*$ (note the difference in $x$-axis range in \autoref{fig:inflow_sfr} versus \autoref{fig:inflow_sfr_mod}), and that the data remain quite close to the model lines. The remaining difference between the theoretical model results for dwarfs and spirals are due to the differences in gas fraction and rotation curve index between the two, and the observations, with one possible exception, are well within the space of models that are plausibly spanned by the gas fraction and rotation curve index range of nearby galaxies. The outlier at the left side of the plot is NGC 2903. (In \autoref{fig:inflow_sfr} this is the point to the at the lowest inflow rate and second lowest star formation rate.) For this galaxy, \citet{schmidt16a} state that their fits are likely unreliable due to complex kinematics driven by a strong bar.

While this comparison is encouraging, the observations from \citet{schmidt16a} cover only a very narrow range of galaxy properties. Stronger tests are clearly warranted. The most obvious target for such comparisons are nearby starburst galaxies, for which our transport+feedback model predicts large inflow rates. These galaxies are nearby enough that one can make high resolution CO or HCN maps from which kinematic information can be extracted. While the kinematics are likely to be complex and thus analysis will be more difficult than it is for quiescent spirals, the predicted signal is also much larger.

\section{Implications for Galaxy Formation}
\label{sec:discussion}

\subsection{Equilibrium Inflow and Star Formation}
\label{ssec:gal_eq}

The model we present here has important implications for the formation of disc galaxies. To explore these further, we begin by changing our perspective. Thus far we have been developing a theory that takes as input the gas content and other ancillary properties of a galaxy, and returns as output the 
inflow rate $\dot{M}$ that is required to maintain pressure and energy equilibrium in a galaxy of given physical parameters, and the star formation rate $\dot{M}_*$ that accompanies this equilibrium configuration. This framing of the problem is appropriate if we are interested in behaviours on timescales short compared to the gas consumption or flow timescales. However, over cosmological timescales it is more natural to think of the inflow rate as given. Gas will fall onto the central galaxy at a rate dictated by cosmological structure formation, recycling of gas ejected at earlier epochs, and processing through the gaseous halo. At least for large galaxies where the gas consumption timescale is shorter than the Hubble time, the galaxy will adapt its structure to be in equilibrium given this inflow rate, a point previously made by \citet{dekel09a}.

We can use this picture to calculate the evolution of galaxies' velocity dispersions. Tidal torque theory suggests that the specific angular momentum of infalling gas will increase with halo mass and cosmic time, and thus gas accreting onto galaxies tends to arrive at their outskirts, where orbital times, and thus star formation timescales, are relatively long. Some of this gas will form stars before gravitational instability moves it inward, but some will be forced to flow inward toward the galactic bulge. This is particularly true at high redshift, when galaxies accrete quickly. We can therefore approximate that the inflow rate must be comparable to the infall rate (including in this rate recycling of gas ejected from the galaxy but not the halo), which in turn dictates the velocity dispersion in the galaxy, as suggested by \citet{genel12b}. In practice, we can calculate this velocity dispersion from \autoref{eq:mdot_steady}, setting $\dot{M} = \dot{M}_{\rm g,acc}$, where $\dot{M}_{\rm g,acc}$ is the gas accretion rate onto the galaxy.

However, from the velocity dispersion and the galaxy rotation curve we can in turn compute the run of gas surface density, and thence the star formation rate. For a simple case of radially-constant $\beta$, $f_{\rm sf}$, and gas fractions, we can compute this analytically using \autoref{eq:sigma_sfr_unresolved}. A more realistic calculation would consider the time and radial variation of the gas fraction and star-forming ISM fraction, and such models have had significant success \citep[e.g.,][]{forbes12a, forbes14a, tonini16a}, but this level of complexity requires semi-analytic solution, and our goal here is qualitative insight. For this reason, we choose to neglect these complications and simply ignore radial- and time-variation. Doing so allows us to compute an equilibrium star formation rate in the disc of the galaxy as a function of a galaxy's inflow rate and rotation curve, which in turn are functions of the halo mass and redshift. We emphasise that this is the star formation rate in the disc; in a simple equilibrium picture, the total star formation rate must equal the accretion rate minus the rate of mass ejection by star formation or black hole feedback. The procedure we have just outlined provides a means of computing the disc star formation rate from the inflow rate and the rotation curve, and with the remaining mass balance coming from a combination of star formation in a bulge and ejection of gas through galactic winds.

\subsection{Cosmological Halo Evolution}
\label{ssec:halo_evol}

To make use of the simple picture outlined above, we must have methods to calculate the halo accretion rate, circular velocity, and orbital period as a function of redshift. First consider the accretion rate. In the interest of simplicity we neglect the contribution from gas recycling, and just attempt to calculate the infall rate of pristine gas. To first order this is determined by the dark matter accretion rate, which in the context of a $\Lambda$CDM cosmology\footnote{Throughout this section we assume a cosmology with $\Omega_m = 0.27$, $\Omega_\Lambda = 0.73$, $h=0.71$, and $\sigma_8 = 0.81$. We use this rather than more recent \textit{Planck} cosmological parameters because we do not have calibrations of the accretion formulae we adopt below for the more recent cosmological parameters. In practice this will make little difference, since our goal in this section is to develop a rough intuitive model rather than perform precision calculations.} can be calculated with the extended Press-Schechter (EPS) formalism, with some additional calibration from simulations. Following \citet{krumholz12d} and \citet{forbes14a}, we adopt the approximate dark matter accretion rates found by \citet{neistein08a} and \citet{bouche10a}:
\begin{equation}
\label{eq:halo_acc}
\dot{M}_{h,12} \approx -\alpha M_{h,12}^{1+\beta} \dot{\omega},
\end{equation}
where $M_{h,12}$ and $\dot{M}_{h,12}$ are the halo mass and accretion rate normalised to $10^{12}$ $M_\odot$, and $\omega$ is the self-similar time variable of the EPS formalism (i.e., $\omega$ is time measured in units of the linear growth time for structure), whose time derivative is well fit by
\begin{equation}
\dot{\omega} \approx -0.0476[1 + z + 0.093(1+z)^{-1.22}]^{2.5}\mbox{ Gyr}^{-1}.
\end{equation}
The functional form of \autoref{eq:halo_acc} follows from the EPS formalism, and the value of $\beta$ follows from the power spectrum of density fluctuations, while $\alpha$ is a free parameter to be calibrated from full dark matter simulations. The \citet{neistein08a} fit, updated to current cosmological parameters, gives $\alpha = 0.628$ and $\beta=0.14$. With these parameters, the accretion rate evaluates numerically to approximately
\begin{equation}
\dot{M}_{h} \approx 39 \, M_{h,12}^{1.1} \left(1+z\right)^{2.2}\,M_\odot\mbox{ yr}^{-1},
\end{equation}
at $z < 1$, with a slightly steeper slope with $1+z$ at higher $z$. In practice, however, rather than use this approximate expression we generate histories of mass versus redshift by direct numerical integration of \autoref{eq:halo_acc}. 

The next step in estimating the baryonic accretion rate is to correct for the efficiency with which baryons penetrate the hot halo of the galaxy. Following \citet{forbes14a}, we compute the gas accretion rate using the model of \citet{faucher-giguere11a},
\begin{eqnarray}
\label{eq:gas_acc}
\dot{M}_{\rm g,acc} & = & \epsilon_{\rm in} f_b \dot{M}_{h} \\
\epsilon_{\rm in} & = & \min\left[\epsilon_0 M_{h,12}^{\beta_{M_h}}(1+z)^{\beta_z}, \epsilon_{\rm max}\right],
\end{eqnarray}
where $f_{\rm b}\approx 0.17$ is the universal baryon fraction, and the parameters $\left(\epsilon_0, \beta_{M_{\rm h}}, \beta_z, \epsilon_{\rm max}\right) = (0.31, -0.25, 0.38, 1)$ are the results of fits to cosmological simulations. This combined with \autoref{eq:halo_acc} enables us to compute the mass accretion rate for an arbitrary halo. The formula for $\epsilon_{\rm in}$ is the result of a fit to the results of a series of SPH simulations run by \citet{faucher-giguere11a}, and it is calibrated to be most accurate for $z > 2$. Despite this we will continue to use it at lower $z$; examination of \citeauthor{faucher-giguere11a}'s figure 9 suggests that it remains reasonably accurate down to $z \approx 0$ for halos below $10^{12}$ $M_\odot$, but that it overestimates the accretion rates for more massive halos at low redshift by a factor of a few.

In addition to the accretion rate, we need the circular velocity and orbital period, or, equivalently, the circular velocity and disk radius. We also follow \citet{forbes14a} in computing the characteristic radius of the disc as
\begin{equation}
r_d \approx 0.035 r_{\rm vir} = 0.035 \cdot 163 \, M_{h,12}^{1/3}(1+z)^{-1}\mbox{ kpc},
\end{equation}
where $r_{\rm vir}$ is the virial radius, the numerical value of 163 kpc is for a halo overdensity of 200, and the coefficient of $0.035$ is roughly consistent with the findings by \citet{kravtsov13a} and \citet{somerville18a} that galaxies have ratios of half-mass to virial radius $r_h/r_{\rm vir} \simeq 0.015 - 0.018$.\footnote{To be precise, since our equilibrium models have gas surface density profiles $\Sigma_{\rm g} \propto 1/r$ for flat rotation curves, implying a mass that scales as $r$, the outer radius should be exactly twice the half-mass radius. Thus our coefficient of 0.035 corresponds to $r_h/r_{\rm vir} = 0.0175$.} For the circular velocity, we note that a \citet{navarro97a} profile has a maximum circular velocity \citep{mo10a}
\begin{eqnarray}
v_{c,\rm max} & \approx & 0.465\sqrt{\frac{c}{\ln(1+c)-c/(1+c)}} v_{\rm vir} \\
v_{\rm vir} & \approx & 117 \, M_{h,12}^{1/3} (1+z)^{1/2} \mbox{ km s}^{-1},
\end{eqnarray}
where $c$ is the halo concentration; the numerical coefficients are again for a halo overdensity of 200, and we adopt $c=10$ as a fiducial value. The true circular velocity will be somewhat larger than this because in the star-forming parts of galaxies the baryons contribute non-negligibly to the gravitational potential. We very roughly adopt a relation
\begin{equation}
v_\phi = \phi_v v_{c,\rm max}
\end{equation}
with $\phi_v = 1.4$, which gives $v_\phi = 200$ km s$^{-1}$ for a $10^{12}$ $M_\odot$ halo at $z=0$. This is a crude approximation, but, as noted above, our goal here is a qualitative toy model, not a precise calculation. The orbital time follows immediately from $r_d$ and $v_\phi$.

\subsection{Model Results and Interpretation}

\begin{figure}
\includegraphics[width=\columnwidth]{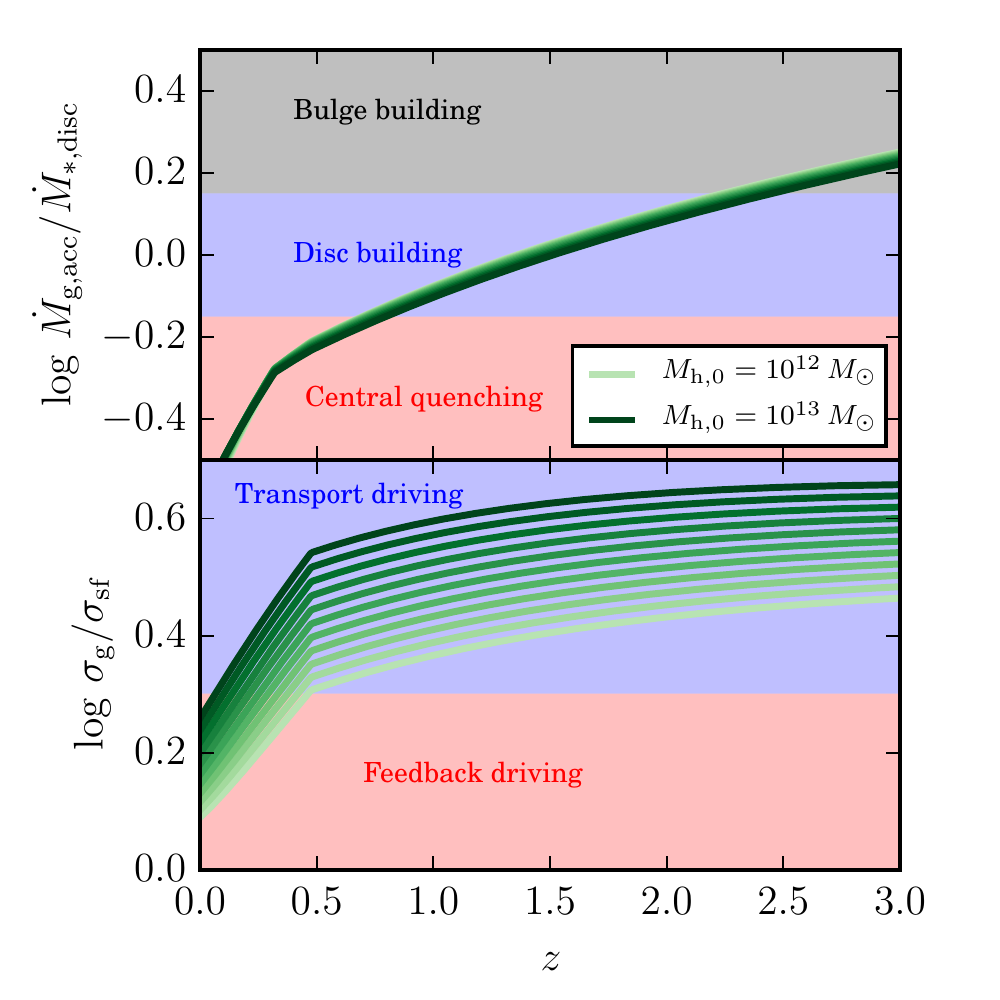}
\caption{
\label{fig:halo_hist}
Evolution of the ratio of gas accretion rate (\autoref{eq:gas_acc}) to disc star formation rate (\autoref{eq:sigma_sfr_unresolved}; top panel) and gas velocity dispersion (\autoref{eq:mdot_steady}) to star formation-supported velocity dispersion (\autoref{eq:sigma_sf}; bottom panel) as a function of redshift. Each line represents the evolutionary path of a particular halo, with the lightest colour (bottom lines in the lower panel) corresponding to a halo with a present-day mass of $M_{h,0} = 10^{12}$ $M_\odot$, and the darkest (top lines in the lower panel) to a halo with a present-day mass of $M_{\rm h,0} = 10^{13}$ $M_\odot$. Intermediate lines are uniformly spaced by 0.1 dex in $M_{\rm h,0}$. The inflection points visible at $z \approx 0.5$ corresponds to where halos switch from star formation occurring mainly in the Toomre regime (at higher $z$) to occurring mainly in the GMC regime (at lower $z$). Shaded regions in the upper panel indicate regimes of bulge building, disc building, and central quenching, and in the lower panel indicate regions of transport-driven versus feedback-driven turbulence; see main text for details.
}
\end{figure}

We now use the formalism of \autoref{ssec:halo_evol} to compute the mass and accretion histories of a range of halos with present-day masses of $M_{h,0} = 10^{12} - 10^{13}$ $M_\odot$. For each one we compute the velocity dispersion $\sigma_{\rm g}$, the velocity dispersion that can be supported by star formation $\sigma_{\rm sf}$, and the disc star formation rate $\dot{M}_{*, \rm disc}$ as outlined in \autoref{ssec:gal_eq}. 

We restrict our attention to halos in this mass range for three reasons. First, halos substantially smaller than this are not observable beyond the local Universe, while those substantially larger host clusters rather than single galaxies. Thus the observational sample beyond $z=0$ is mostly limited to this mass range. Second, it is not clear if the equilibrium assumption is applicable in halos smaller than $\sim 10^{12}$ $M_\odot$. These host dwarf galaxies, and the gas consumption timescale in modern-day dwarfs is generally comparable to or longer than the Hubble time \citep[e.g.,][]{bolatto11a, hunter12a, jameson16a}. This might also be the case at higher redshift, but this is uncertain because the gas consumption timescale depends on both the star formation rate and the rate at which star formation drives gas out via galactic winds, and the latter is highly uncertain for low-mass galaxies beyond the local universe \citep{forbes14b}. Limiting our attention to halos above $10^{12}$ $M_\odot$ avoids this issue. A related point is that the mass loading factor for these halos is unlikely to be $\gtrsim 1$, so we need not adopt a complex model to treat this phenomenon either. Third and finally, to evaluate our models we require values of $f_{\rm sf}$ and $\sigma_{\rm th}$, and these depend on the molecular fraction in the ISM. The dependence of this fraction on halo mass and redshift is highly complex and substantially uncertain \citep[e.g.,][]{obreschkow09a, fu10a, lagos11a, krumholz12d, forbes14a}. However, we expect that the molecular fraction will be smallest in small halos at low redshift, since these combine low metallicity and low gas surface density. At $z=0$, we observe that halos of mass $10^{12}$ $M_\odot$ halos (Milky Way-sized) host galaxies with $f_{\rm H_2}\sim 0.5$ within the scale radius. By restricting our attention to halos with present-day masses above this limit, we stay in the part of parameter space where the ISM is at least marginally molecule-dominated, and thus we can adopt $f_{\rm sf} \approx 1$ and $\sigma_{\rm th} \approx 0.2$ km s$^{-1}$ without making too large an error. In contrast, $10^{11}$ $M_\odot$ halos (LMC-sized) have present-day molecular fractions $\sim 0.1$ \citep[e.g.,][]{jameson16a}. Treating them would require that we adopt a model for the time evolution of the molecule fraction, which is beyond the scope of this paper.

\subsubsection{Star formation}

We are interested in two diagnostic ratios from these models, which we plot in the upper and lower panels of \autoref{fig:halo_hist}. The first of these is $\dot{M}_{\rm g, acc}/\dot{M}_{*, \rm disc}$, the ratio between the rate of accretion onto and then through the disc and the rate at which that accretion flow should convert into stars as it moves through the disc; \citet{dekel14a} describe this quantity as the ``wetness factor". If $\dot{M}_{\rm g,acc} \gg \dot{M}_{*,\rm disc}$, then the rate of mass flow onto the disc from the IGM, and through it toward the galactic centre, greatly exceeds the rate at which we expect that flow to convert into stars. Consequently, the majority of the flow will not be converted to stars before it reaches the galactic centre. It may still be ejected in outflows, but unless all of it is lost in this fashion, a substantial mass flux will reach the bulge region. Consequently, an era when $\dot{M}_{\rm g,acc} \gg \dot{M}_{*,\rm disc}$ should correspond to an era when galaxies are building up their bulges.

Now suppose the reverse holds, $\dot{M}_{\rm g,acc} \ll \dot{M}_{*,\rm disc}$. Taken literally this would mean that the star formation rate exceeds the gas accretion rate. Of course such a configuration cannot represent a steady state in which the rate of gas flow through the galaxy matches the rate of gas accretion into it, which violates the assumption we made in deriving $\dot{M}_{*,\rm disc}$. To understand what occurs in this regime, it is helpful to consider what happens as a galaxy evolves, with our intuition guided by the results of more detailed time-dependent models \citep[e.g.,][]{forbes14a}. From  \autoref{fig:halo_hist} we see that at early times halos have $\dot{M}_{\rm g, acc}/\dot{M}_{*, \rm disc} \gg 1$, and that this ratio gradually decreases to $\sim 1$. When this ratio is $\sim 1$, gas only barely reaches the galactic centre before the last of it is consumed and turned into stars; all star formation therefore occurs in the disc. As the gas supply tapers off over cosmic time and the ratio tries to drop even further, the gas supply is insufficient to keep up with the rate of consumption into stars. The equilibrium between supply and consumption is easiest to maintain in the outer parts of galaxies, both because this is where the majority of the gas lands, and because this is where the star formation rate per unit area is smallest. Thus the failure of equilibrium is likely to occur inside out: less and less of the gas that is accreting onto the galaxy will be able to reach the centre before transforming into stars. This reduction in central gas surface density in turn reduces $\dot{M}_{*,\rm disc}$ compared to the value we have computed under the assumption of constant radial mass flux, thereby maintaining $\dot{M}_{\rm g,acc} \sim \dot{M}_{*,\rm disc}$. The price of this balance is that the centre of the galaxy ceases star formation, and thus quenches.

Examining the upper panels of \autoref{fig:halo_hist}, we see halos show a clear progression from bulge building at high-redshift to disc building at intermediate redshift, to central quenching near the present day. We should not put too much weight on the exact redshift range over which this transition occurs, particularly since our fitting formula for gas accretion onto halos likely overestimates the rate of cold gas flow at lower redshift. Nonetheless, qualitatively this progression is a natural explanation for a commonly-observed phenomenon: galaxies that transition from the blue, star-forming cloud to the quenched, red sequence do so by ceasing their star formation from the inside out, after a central stellar bulge builds up \citep[e.g.,][]{fang12a, fang13a, cheung12a, genzel14a, nelson16a, belfiore17a}. Models that include radial transport of gas via gravitational instability are able to reproduce this qualitative behaviour \citep[e.g.,][]{forbes12a, cacciato12a, forbes14a, tonini16a, stevens16a}, but the analytic model we develop here allows a particularly simple and straightforward explanation for both the inside-out quenching phenomenon and the redshift at which it occurs. 

However, we end this section by cautioning that this simple steady state, quasi-equilibrium picture almost certainly misses some of the complications that occur in real cosmological galaxy formation. At least for halos at and slightly above the more massive end of the range we consider, which tend to quench at $z\sim 2$, both observations \citep{barro13a, barro16a, barro17a, tacchella15a, tacchella18a} and simulations \citep{zolotov15a, tacchella16a, tacchella16b} suggest that galaxies pass through a phase of ``compaction", where the central gas surface density is driven to very high values, before finally quenching. Such compaction events are not captured in our simple steady state model, which may be more applicable to galaxies in less massive halos such as those that host the Milky Way.

\subsubsection{Turbulence driving}

The other diagnostic ratio of interest is $\sigma_{\rm g} / \sigma_{\rm sf}$. Recall that $1-\sigma_{\rm sf}/\sigma_{\rm g}$ is the fraction of the energy required to maintain the turbulence that comes from star formation feedback (\autoref{eq:sigma_sf}). Thus $\sigma_{\rm g} = 2 \sigma_{\rm sf}$ corresponds to star formation feedback and transport (gravity) contributing equally to the turbulence. In the lower panel of \autoref{fig:halo_hist}, we show the time evolution of $\sigma_{\rm g} / \sigma_{\rm sf}$, with the regions where transport driving and feedback driving dominate the energy budget highlighted. Note that, under our simplifying assumption that the inflow rate is always non-zero, we cannot ever reach $\sigma_{\rm g} = \sigma_{\rm sf}$, corresponding to the point where transport driving ceases completely.

The most obvious trend in \autoref{fig:halo_hist} is that more massive halos are further into the transport-driving regime, and spend more of their evolutionary history there. Massive galaxies, by virtue of their large accretion rates, tend to have high gas surface densities that require high velocity dispersions to maintain. Such velocity dispersions can only be maintained by gravitational power. The second clear trend is that $\sigma_{\rm g}/\sigma_{\rm sf}$ drops as we approach $z=0$. This is driven partly by a drop off in cosmological accretion rates, which produce less gas-rich discs that require lower values of $\sigma_{\rm g}$ to remain marginally stable. It is also partly by galaxies transitioning from the Toomre regime to the GMC regime in their star formation, which puts a ceiling on the depletion time and thus a floor on $\sigma_{\rm sf}$. This effect is visible as the downturn in $\sigma_{\rm g}/\sigma_{\rm sf}$ below $z\approx 0.5$.

Thus the qualitative picture to which we come is that the transition from transport-driven turbulence to feedback-driven turbulence depends primarily on galaxy mass, and secondarily on redshift, with transport-driving dominating at high mass and high-$z$.

\subsection{Limits of model applicability}
\label{ssec:limitations}

We conclude this discussion by considering to what types of galaxies, or within which parts of galaxies, our model applies. As noted in \autoref{sec:model}, our model assumes that the gas inflow rate is able to self-adjust to maintain marginal gravitational stability. The numerous simulations discussed in \autoref{ssec:theory} leave little doubt that this self-adjustment works in one direction: if the gas becomes gravitationally unstable, it will develop non-axisymmetric structures that exert torques and drive a net inflow, thereby raising the velocity dispersion and pushing the system back toward stability. However, our model also assumes the converse, that in galaxies that are gravitationally stable there will not be net radial transport. There are clearly locations where this is not true, such as the inner few kpc of the Milky Way. In this region the gas fraction is so small that the gas effectively acts like a passive tracer moving in the fixed stellar potential. Both the gas and the combined gas-star disc are Toomre-stable, but the stars are arranged in a bar that, depending on the galactocentric radius, either forces gas to shock by preventing it from flowing on non-intersecting orbits \citep[e.g.,][]{binney91a, sormani15a} or drives acoustic instabilities that are unrelated to self-gravity \citep[e.g.,][]{montenegro99a, krumholz15d, krumholz17a}. Moreover, these effects can drive gas outward as well as in. In any region where such effects are dominant, our model is not applicable.

However, this is not a major limitation because regions such as the Milky Way centre contribute very little to the total gas mass or star formation budget of the Universe. The Milky Way's centre is gas-depleted to an extent that is unusual even among local spirals \citep{bigiel12a}, and is likely related to it being a ``green valley" galaxy on the verge of quenching \citep{bland-hawthorn16b}. Even including the central molecular zone, the central few kpc of the Galaxy contain $\sim 10\%$ of its star formation or ISM mass \citep[e.g.,][]{kruijssen14b}. Indeed, this gas-poverty is likely the reason that there is a strong bar, since simulations show that bar formation does not take place until the gas fraction drops to well under 10\% \citep{athanassoula13a}. 

Beyond the Milky Way, the majority of gas in nearby spirals lies in regions where the gas and stellar surface densities and velocity dispersions are such that gas and stars are strongly coupled \citep[their Figure 5]{romeo17a}. Estimates of the fraction of galaxies that contain stellar bars at all vary from $\sim 20\%$ \citep[e.g.,][]{melvin14a, cervantes-sodi17a} to $\sim 60\%$ \citep[e.g.,][]{erwin17a}, with the higher figures largely coming from surveys capable of detecting bars below $\sim 1$ kpc in size. Thus large bars that could conceivably affect the dynamics of the majority of the ISM mass or star formation in a galaxy appear to be rare even at $z=0$, as one might expect since even a strongly-barred galaxy like the Milky Way has little of its gas or star formation in the region where the bar dominates the dynamics. Though there is significant debate in the literature about whether the bar fraction declines with redshift (e.g., \citealt{melvin14a} versus \citealt{erwin17a}), there are strong theoretical reasons to believe that bars were less prominent in the past, both due to the higher gas fractions found at $z>0$  \citep[e.g.,][]{tacconi13a} and the fact that bars take time to grow. Thus if barred regions do not dominate the ISM mass budget at $z=0$, they should be even less important in the early Universe. In summary, our model should apply to most of the ISM and most star-forming regions over most of cosmic time. We do, however, caution that it should not be applied to extremely gas-poor, bar-dominated regions like the Milky Way centre. These are better described by models that treat the stellar potential as decoupled from the gas \citep[e.g.,][]{binney91a, sormani15a, krumholz17a}.

\section{Summary}
\label{sec:summary}

We present a new model for the gas in galactic discs, based on a few simple physical premises. We propose that galactic discs maintain a state of vertical hydrostatic equilibrium, marginal gravitational stability, and balance between dissipation of turbulence and injection of turbulent energy by star formation feedback and radial transport. The inclusion of both radial transport and feedback as potential energy sources is the primary new feature of our model, and despite the apparent simplicity of this addition, it yields a dramatic improvement in both predictive power and agreement with observation compared to simpler equilibrium models.

We find that star formation alone is able to maintain a velocity dispersion of $\sigma_{\rm sf} \approx 6 - 10$ km s$^{-1}$, with the exact value depending on the gas fraction, the thermal velocity dispersion, and the fraction of the interstellar medium in the star-forming molecular phase. In galaxies where the gas surface density is low, this is sufficient to maintain energy and hydrostatic balance, and there is no net radial flow. However, in many observed galaxies this velocity dispersion is insufficient to keep the gas in a state of marginal gravitational stability. In this case, the instability produces spiral structures and clumps that exert non-axisymmetric torques, leading to a net mass flow inward. The inflow releases gravitational potential energy, which manifests as non-circular, turbulent motions in the transported gas. The fraction of the turbulent power that originates from this process rather than from star formation feedback is $1 - \sigma_{\rm sf}/\sigma_{\rm g}$, where $\sigma_{\rm g}$ is the gas velocity dispersion. This fraction is small in quiescent star-forming galaxies at the present cosmic epoch, but is larger in both starbursts and high-redshift galaxies.

The model we derive from this simple picture shows excellent agreement with a range of observational diagnostics, including the star formation law for both resolved and unresolved systems and the relationship between galaxies' star formation rates and velocity dispersions. It is also consistent with the limited data available on the observed rates of radial inflow in nearby galaxies. The agreement holds across a wide range of galaxy types and masses, from nearby dwarfs with star formation rates $\lesssim 0.1$ $M_\odot$ yr$^{-1}$, to starbursts and high-redshift discs with star formation rates $\gtrsim 100$ $M_\odot$ yr$^{-1}$. We also predict that high gas inflow rates should be measurable in nearby starburst galaxies, whose kinematics have yet to be analysed for inflow. In contrast, we show that models that neglect either radial transport or star formation feedback fail at either high or low star formation rate, or in some cases both.

Our model provides a natural explanation for the cosmic epochs at which galaxies build up bulges and discs, and at which they quench. At high redshift, galaxies' mass transport rates naturally exceed their star formation rates; this is a natural consequence of the high velocity dispersions found in high-redshift galaxies, and the stronger scaling between velocity dispersion and transport rate ($\dot{M} \propto \sigma_{\rm g}^3$) than between velocity dispersion and star formation rate ($\dot{M}_* \propto \sigma_{\rm g}$). As a result, they tend to move mass inward toward a bulge. As accretion rates decline as the density of the universe diminishes, transport rates decline as well, and do so faster than star formation rates. This leads to a configuration where most star formation occurs in galaxies' discs. Finally, once the star formation rate is smaller than the mass transport rate, gas does not reach galaxy centres at all, and the centres quench, explaining the common observation that quenching tends to occur inside out.

In future work we plan to apply this model to radially-dependent models of galaxy formation, such as those of \citet{forbes12a} and \citet{forbes14a}. Such an application promises to yield new insights into the origin of the radial structure of galactic discs, and the evolution of this structure over cosmic time. We also plan to test the model against cosmological simulations, where inflow rates are determined directly from the hydrodynamics (Burkhart et al.~2018, in preparation).

\section*{Acknowledgements}

MRK acknowledges support from the Australian Research Council's \textit{Discovery Projects} funding scheme (project DP160100695). BB is supported by the NASA Einstein Postdoctoral Fellowship and an Institute for Theory and Computation Fellowship. JCF is supported by an Institute for Theory and Computation Fellowship. We thank S.~Oh for helpful discussions of bar surveys, and the referee, A.~Dekel, for helpful comments.

\bibliographystyle{mn2e}
\bibliography{refs}

\begin{appendix}

\section{Choice of $\alpha_{\rm CO}$}
\label{app:alpha_CO}

Our sample of galaxies includes starburst and high-redshift systems, and for these galaxies there is significant uncertainty about the choice of $\alpha_{\rm CO}$, the conversion factor between CO luminosity and gas surface density \citep{bolatto13a}. Throughout this paper, we have chosen to adopt the convention for $\alpha_{\rm CO}$ recommended by \citet{daddi10a}: $\alpha_{\rm CO} = 4.6$ $M_\odot$ (K km s$^{-1}$ pc$^{-2}$) for local spirals, $\alpha_{\rm CO} = 3.6$ $M_\odot$ (K km s$^{-1}$ pc$^{-2}$) for high-redshift discs, and $\alpha_{\rm CO} = 0.8$ $M_\odot$ (K km s$^{-1}$ pc$^{-2}$) for starbursts / ULIRGs / sub-mm galaxies at all redshifts.

We note that, in contrast, \citet{faucher-giguere13a} adopt a theoretical $\alpha_{\rm CO}$ from \citet{narayanan12a}. The effects of this choice are modest for most galaxies, but are significant for high-redshift star-forming disc galaxies, e.g., the PHIBSS sample of \citet{tacconi13a}. Using the \citet{narayanan12a} $\alpha_{\rm CO}$ for these objects gives them a ULIRG-like $\alpha_{\rm CO}$ rather than a Milky Way-like one favoured by the observers, and that we have adopted. This in turn shifts these objects to lower $\Sigma_{\rm g}$ by a factor of $\sim 5$, and produces a steepening of the best-fit slope to $\approx 1.7$, compared to $\approx 1.4$ for the conventional $\alpha_{\rm CO}$; we refrain from plotting the data and fit here, since such a plot is shown in Figure 3 of \citet{thompson16a}.

The question of the correct $\alpha_{\rm CO}$ for disc galaxies on the star-forming main sequence at high redshift has recently be reinvestigated by \citet{genzel15a} using dust observations, which provide an independent means of measuring the gas mass. They find that the conventional $\alpha_{\rm CO}$ (i.e., one similar to that of the Milky Way) is a better match to their data than a ULIRG-like $\alpha_{\rm CO}$. We therefore retain the conventional $\alpha_{\rm CO}$ in this work.

\section{Velocity Dispersion Data}
\label{app:sigma_sfr_data}

Here we provide details on our handling of the observations of velocity dispersion $\sigma$ versus star formation rate $\dot{M}_*$, as discussed in \autoref{ssec:sfr_vdisp}. A general issue that arises when combining multiple data sets is that we are interested in the velocity dispersion of the cool atomic or molecular ISM, but much of the data available (and all of the high-redshift data) are H$\alpha$ measurements, which are likely dominated by gas in H~\textsc{ii} regions. Such gas has a thermal velocity dispersion of $\approx 10$ km s$^{-1}$, and cometary expansion of photoionised gas adds a comparable non-thermal component on top of this. The effects of this are evident if one compares the velocity dispersions obtained from H~\textsc{i} observations to those obtained by H$\alpha$ measurements in galaxies of similar, low star formation rates; the H$\alpha$ velocity dispersions are a factor of $\sim 2$ larger. Since the H$\alpha$ excess is larger than $\sigma_{\rm sf}$, we must remove it in order to make meaningful comparisons. Following \citet{krumholz16a}, we adopt a combined thermal plus non-thermal velocity dispersion of $\sigma_{\rm H~\textsc{ii}} = 15$ km s$^{-1}$ for H~\textsc{ii} region gas, and subtract this in quadrature where needed. However, even after this correction the H$\alpha$ velocity dispersions in low star formation rate galaxies remain systematically larger than the values obtained by H~\textsc{i} or molecular lines. There is no comparable discrepancy between H$\alpha$ and molecules at high star formation rates, so the most likely explanation for the offset at low star formation rates is contamination by motions within H~\textsc{ii} regions, which is a more severe problem when the overall velocity dispersion is lower.

Details on individual data sets follow, and our data compilation is available from \url{https://bitbucket.org/krumholz/kbfc17}.

\textbf{GHASP.} The GHASP survey include H$\alpha$ measurements for a range of local galaxies; values of $\sigma$ and $\dot{M}_*$ are reported in \citet{epinat08a}; we retrieved the data from the VizieR database entry associated with the paper. The velocity dispersions $\sigma_{\rm g}$ we plot are listed as $\sigma_{\rm res}$ in their tables. Despite the fact that these are H$\alpha$ measurements, the fitting method essentially measures the dispersion of velocity centroids inside each fitted ring, and thus is insensitive to the broadening of the line. For this reason, we do not subtract $\sigma_{\rm H~\textsc{ii}}$ from these data points.

\textbf{\citet{epinat09a}.} This paper reports H$\alpha$ measurements for galaxies at $1.2 < z < 1.6$. The quantity we plot as $\sigma_{\rm g}$ is the local mean velocity dispersion $\sigma_0$ given in their Table 5, with $\sigma_{\rm H~\textsc{ii}}$ subtracted in quadrature.

\textbf{\citet{law09a}.} This paper reports H$\alpha$ and [O~\textsc{iii}] observations of galaxies at $z\sim 2 - 3.3$. For our star formation rate estimate, we use the values given in the paper based on nebular emission (their $\mbox{SFR}_{\rm neb}$), and the velocity dispersion we use is their $\sigma_{\rm mean}$, with $\sigma_{\rm H~\textsc{ii}}$ subtracted in quadrature. While the correction factor should be somewhat smaller for [O~\textsc{iii}] than for H$\alpha$ due to the smaller thermal broadening of the heavy ion, the velocity dispersions reported are so large that the quadrature subtraction has not noticeable effect in any event. For galaxies where more than one value of $\sigma_{\rm mean}$  is given, we plot an average of the values listed.

\textbf{\citet{jones10a}.} This paper reports H$\alpha$ observations of gravitationally-lensed galaxies at $z\sim 2-3$. We take our star formation rates and velocity dispersions directly from their tables, subtracting $\sigma_{\rm H~\textsc{ii}}$ in quadrature.

\textbf{\citet{lemoine-busserolle10a}.} This study reports H$\alpha$ star formation rates and velocity dispersions from galaxies at $z\sim 3$. The value we plot as $\sigma_{\rm g}$ is their $\sigma_{\rm mean}$, with $\sigma_{\rm H~\textsc{ii}}$ subtracted in quadrature.

\textbf{WiggleZ.} We plot a sample of H$\alpha$ observations of galaxies from the WiggleZ survey as repoted by \citet{wisnioski11a}. We convert their reported H$\alpha$ luminosities to star formation rates using the conversion given by \citet{kennicutt12a}, and we use their reported $\sigma_{\rm mean}$ as the gas velocity dispersion $\sigma_{\rm g}$ after subtracting $\sigma_{\rm H~\textsc{ii}}$ in quadrature.

\textbf{THINGS.} We use the global H~\textsc{i} velocity dispersions measured in local galaxies by the the THINGS survey, as reported by \citet{ianjamasimanana12a}. Our $\sigma_{\rm g}$ is their single Gaussian fit value. We take corresponding star formation rates from \citet{leroy08a} if they are given there, or from the literature compilation of \citet{walter08a} if not.

\textbf{\citet{stilp13a}.} This paper reports H~\textsc{i} measurements of local dwarf galaxies. For $\sigma_{\rm g}$ we use their $\sigma_{\rm central}$, the velocity dispersion of their central Gaussian component.

\textbf{DYNAMO.} The DYNAMO survey reports H$\alpha$ measurements of local galaxies selected to be have conditions similar to those commonly seen at high-$z$. Kinematic data are reported in \citet{green14a}, and we retrieve the quantities we plot from the VizieR entry associated with the paper. We use their H$\alpha$-estimated star formation rates, and their estimated $\sigma$ as the velocity dispersion, after subtracting $\sigma_{\rm H~\textsc{ii}}$ in quadrature.

\textbf{\citet{moiseev15a}.} This paper reports a survey of H$\alpha$ emission from local dwarf galaxies. We derive star formation rates for this sample by converting their reported H$\alpha$ luminosities using the conversion of \citet{kennicutt12a}. For the velocity dispersion, the values reported in the paper have already been corrected for an assumed thermal broadening of 9.1 km s$^{-1}$. For consistency with the other data sets, we remove this correction by adding 9.1 km s$^{-1}$ in quadrature, then subtracting $\sigma_{\rm H~\textsc{ii}}$ in quadrature.

\textbf{\citet{varidel16a}.} This is a sample of local analogs of high-$z$ galaxies similar to DYNAMO. For these galaxies we use the star formation rates derived from line emission, and for velocity dispersion we use the flux-weighted mean ($\sigma_m$ in \citeauthor{varidel16a}'s notation).

\textbf{\citet{di-teodoro16a}.} This is an analysis of H$\alpha$ observations of galaxies from the KMOS survey \citep{wisnioski15a} using a novel kinematic fitting method. We use the star formation rates and velocity dispersions taken from their Table 1, subtracting $\sigma_{\rm H~\textsc{ii}}$ in quadrature.

\textbf{SINS.} The SINS-KMOS-3D survey \citep{wisnioski15a} includes H$\alpha$ measurements for 248 $z\sim 1-3$ galaxies, as reported by \citet{wuyts16a}. The velocity dispersions and star formation rates were provided by R.~Genzel (2017, priv.~comm.). The values we plot have $\sigma_{\rm H~\textsc{ii}}$ subtracted in quadrature.

\textbf{KROSS.} The KMOS Redshift One Spectroscopic Survey (KROSS; \citealt{stott16a}) is a spectroscopic survey of redshift one galaxies using KMOS on the VLT. \citet{johnson17a} provides measurements of velocity dispersion based on the H$\alpha$ line in 472 galaxies from the survey, along with star formation rates for the same galaxies. The values we plot have $\sigma_{\rm H~\textsc{ii}}$ subtracted in quadrature.

\textbf{ULIRGs.} We have compiled a new sample of measured velocity dispersions in local ULIRGs, drawn from the literature. Velocity dispersions for all galaxies except NGC 6240 are based on CO measurements; NGC 6240 is based on HCN. The data are given in \autoref{tab:ulirg_data}.

\begin{table}
\begin{tabular}{c@{$\quad$}cccc}
\hline
Name & $\dot{M}_*$ [$M_\odot$ yr$^{-1}$] & $\sigma_{\rm g}$ & $\sigma_{\rm g}$ Ref & $\dot{M}_*$ Ref \\
\hline
Arp 220 West & 120 & 128 & (1) & (1) \\
Arp 220 East & 120 & 61 & (1) & (1) \\
NGC 6240 & 70 & 160 & (2) & (2) \\
Mrk 231 & 176 & 60 & (3) & (4) \\
VII Zw 31 & 66 & 21 & (3) & (5) \\
IRAS 10565+2448 & 163 & 29 & (3) & (4) \\
Arp 193 & 66 & 29 & (3) & (5) \\
IRAS 17208-0014 & 428 & 107 & (3) & (4) \\
IRAS 23365+3604 & 102 & 71 & (3) & (4) \\
\hline
\end{tabular}
\caption{
\label{tab:ulirg_data}
Compilation of ULIRG data. References are as follows: (1) \citet{scoville17a}; (2) \citet{scoville15a}; (3) \citet{downes98a}; (4) \citet{veilleux09a}; (5) \citet{sanders03a}. For galaxies where the star formation rates are from (4) or (5), we have converted the reported total IR luminosity to a star formation rate using the conversion of \citet{kennicutt12a}. For IR data from \citet{veilleux09a}, we subtract off the AGN contribution using their estimates. For IR data from \citet{sanders03a}, we adopt an AGN fraction of 50\%.
}
\end{table}

\end{appendix}

\end{document}